\documentclass{jpsj3}
\usepackage{graphicx,color}

\def\diff{\mathrm d}
\newcommand{\etal}{{\it et al.}\ }

\title{
Second-order Perturbation Formula for Magnetocrystalline Anisotropy using Orbital Angular Momentum Matrix
}
\author{Taichi Kosugi$^{1,2}$, Takashi Miyake$^1$, and Shoji Ishibashi$^1$}

\inst{
$^1$Nanosystem Research Institute (NRI) ``RICS'', AIST, 1-1-1 Umezono, Tsukuba 305-8568, Japan \\
$^2$RIKEN, Advanced Institute for Computational Science, 7-1-26, Minatojima-minami-machi, Chuo-ku, Kobe, Hyogo 650-0047, Japan
}

\abst{
We derive a second-order perturbation formula for an electronic system subject to spin-orbit interactions (SOI).
The energy correction originates in the spin-conserving and the spin-flip transitions.
The former are represented by the orbital angular momentum (OAM) acquired via the SOI.
The latter come from the quantum fluctuation effect.
By using our formula,
we examine the relativistic electronic structures of a $d$ orbital chain and $L1_{0}$ alloys.
The appearance of OAM in the chain is understood by using a parabolic-bands model and the exact expressions of the single-particle states.
The total energy is found to be accurately reproduced by the formula.
The self-consistent fully relativistic first-principles calculations based on the density functional theory are performed for five $L1_0$ alloys.
It is found that the formula reproduces qualitatively the behavior of their exact magnetocrystalline anisotropy (MCA) energies.
While the MCA of FePt, CoPt, and FePd originates in the spin-conserving transitions,
that in MnAl and MnGa originates in the spin-flip contributions.
For FePt, CoPt, and FePd,
the tendency of the MCA energy with the variation in the lattice constant obeys basically that of the spin-flip contributions.
These results indicate that not only the anisotropy of OAM,
but also that of spin-flip contributions must be taken into account for the understanding of the MCA of the $L1_0$ alloys.
}

\kword{magnetocrystalline anisotropy, spin-orbit interaction, first-principles calculation}

\begin{document}
\maketitle

\section{Introduction}

The anisotropy of magnetic properties has been attracting much attention via the recent development of technological applications.
The magnetic anisotropy is seen in materials of various geometries and dimensionalities such as bulks, nanoparticles, surfaces, and wires.
Among them, the anisotropic properties in periodic systems are called the magnetocrystalline anisotropy (MCA), which are described by the classical magnetic dipolar interactions and the electron-ion interactions.
Such a classical dipolar interaction is known to originate from the relativistic quantum mechanical two-electron interaction, called the Breit interaction\cite{bib:Breit}.
It gives rise to not only the effective dipole-dipole interactions, but also the quadrupole-quadrupole ones contributing to the MCA\cite{bib:2905,bib:2915,bib:2195}.
We do not, however, take them into account in the present study since these electron-electron contributions are much smaller than the electron-ion interactions in general.
We denote the MCA coming from the electron-ion interactions simply by the MCA in the present study.
It is widely accepted that the physical origin of the MCA is the anisotropy of the orbital angular momentum (OAM) caused by the spin-orbit interaction (SOI)\cite{bib:2905, bib:2906}.

For electronic structure calculations based on the density functional theory (DFT)\cite{bib:76, bib:77},
the force theorem\cite{bib:Force_Theorem, bib:2196} ensures that MCA energy can be calculated only from the perturbed energy eigenvalues for different spin configurations.
This theorem has been used for the calculations of MCA energy by introducing the SOI as perturbation into the Kohn-Sham Hamiltonian. 
The state-tracking method\cite{bib:state_tracking} as a way for elaboration of the calculation of MCA energy using the force theorem has been proposed.
While the force theorem is used primarily for obtaining the MCA energy in a perturbative DFT calculation,
our formula provided below should be used for an analysis of the results for which the MCA energy has been obtained in self-consistent fully relativistic (FR) DFT calculations.

Bruno\cite{bib:2143} derived a formula for the energy correction based on the second-order perturbation theory for an electronic system in the presence of SOI.
His formula expresses explicitly the connection between the OAM induced by the SOI and the MCA in a ferromagnet.
It is often used for the analyses of the results obtained in model and first-principles calculations\cite{bib:1421, bib:2608, bib:2623, bib:2795}.
An extension of the Bruno's formula containing the spin-flip contributions in an approximated way has been proposed\cite{bib:2624}.

In the present study,
we first derive a second-order perturbation formula for the correction to the energy eigenvalue of a many-body electronic state under SOI.
The formula is reduced to the Bruno's formula in a certain limit.
We then examine the appearance of net OAM in a periodic system via SOI by using a parabolic-bands model.
As applications of the formula,
we examine the relativistic electronic structures of two examples, a $d$ orbital chain by and $L1_{0}$ alloys.
The appearance of the net OAM and the energy correction for the $d$ orbital chain subject to SOI is analyzed
by performing tight-binding calculations.
We pay particular attention to the order of perturbation for the chain.
The origin and the behavior of the $L1_0$ alloys are examined by performing self-consistent FR DFT calculations.
We focus on the difference in OAM and MCA energy between the alloys.

\section{Theory}

\subsection{Perturbation Hamiltonian}

Let 
\begin{gather}
	H = H_0 + H_{\mathrm{SO}}
\end{gather}
be the many-body Hamiltonian of an electronic system.
We assume that the spatial part of the many-body ground state $| \Psi_0 \rangle$ for the unperturbed Hamiltonian $H_0$ is nondegenerate.
This assumption ensures that the wave function of the ground state is the same as its complex conjugate apart from a phase factor.
Since the OAM operator $\boldsymbol{L}$ in spatial representation is purely imaginary,
the OAM for the ground state in this case vanishes: $\langle \Psi_0 | \boldsymbol{L} | \Psi_0 \rangle = 0$, well known as the quench of OAM.
The unperturbed state changes when the SOI represented by the perturbation Hamiltonian $H_\mathrm{SO}$ is turned on.
We assume that all the electron spins in $| \Psi_0 \rangle$ are collinear.
Since the first-order energy correction for the ground state vanishes due to the quench of OAM,
the energy correction to the many-body state within the second-order perturbation can be calculated by using only the perturbed ground state:
\begin{gather}
	\delta E_0
		= \frac{1}{2} \langle \Psi | H_\mathrm{SO} | \Psi \rangle 
	.
	\label{delta_E_0}
\end{gather}

When we adopt the single-particles picture for a periodic system, however,
it should be noted that a nondegenerate single-particle wave function with a nonzero wave vector $\boldsymbol{k}$ can have an OAM even when the SOI is absent.
It is because that its complex conjugate has a wave vector $-\boldsymbol{k}$, which in general does not ensure its coincidence with the wave function with $\boldsymbol{k}$.
This fact allows each single-particle state to undergo the first-order correction of the energy eigenvalue due to the SOI.

When the valence electrons in the vicinity of each ion are spin-polarized,
the potentials they feel depend on their spin directions (parallel or antiparallel to the quantization axis $\boldsymbol{n}$) due to the exchange interactions even if the SOI is absent.
With the SOI in the crystal turned on,
its strength thus differ for the spin direction of each electron 
since the SOI originally comes from the gradient of an electrostatic potential.\cite{bib:Sakurai_Advanced}
To describe such a situation,
we assume that the SOI is the sum of the contributions from the individual atoms in the crystal
and the perturbation Hamiltonian takes the following generic form:
\begin{gather}
	H_{\mathrm{SO}} = \sum_{ \mu} Q_\mu   \boldsymbol{L}_\mu \cdot \boldsymbol{S} Q_\mu
	,
	\label{def_H_SOI}
\end{gather}
where the hermitian operator
\begin{gather}
	Q_\mu (\boldsymbol{n}) \equiv
		\sqrt{\xi_\mu^\uparrow} P_\uparrow + \sqrt{\xi_\mu^\downarrow} P_\downarrow
	\label{def_Q_SOI}
\end{gather}
has been introduced so that the electrons with a different spin direction feel a different strength of the SOI around the atom $\mu$.
The OAM operator $\boldsymbol{L}_\mu$ is effective only in the vicinity of the atom $\mu$.
The spin operator $\boldsymbol{S} = \boldsymbol{\sigma}/2$ is the half of the Pauli matrix.
$P_\uparrow = | \boldsymbol{n} \rangle \langle \boldsymbol{n} |$ is the spin projection operator for the spin-up electrons for the quantization axis $\boldsymbol{n}$,
while $P_\downarrow =  | -\boldsymbol{n} \rangle \langle -\boldsymbol{n} |$ is that for the spin-down electrons.
The two-component spinors $| \boldsymbol{n} \rangle$ and $| -\boldsymbol{n} \rangle$  represent the spin-up and the spin-down states, respectively,
whose expectation values of the spin operator are $\langle \pm \boldsymbol{n} | \boldsymbol{S} | \pm \boldsymbol{n} \rangle = \pm \boldsymbol{n}/2$.
$\xi_\mu^\uparrow$ ($\xi_\mu^\downarrow$) is the strength of the SOI for the spin-up (spin-down) valence electrons.
Sakuma\cite{bib:2224,bib:2244} calculated the strengths of SOI for each direction of electron spins for an analysis of MCA.
If we set $\xi_\mu^\uparrow = \xi_\mu^\downarrow  \equiv \xi_\mu$,
the perturbation Hamiltonian becomes of the well known $\boldsymbol{n}$-independent form, $H_{\mathrm{SO}} = \sum_{ \mu} \xi_\mu  \boldsymbol{L}_\mu \cdot \boldsymbol{S}$.

\subsection{Derivation of Second-order Perturbation Formula}

We decompose the OAM operator around the atom $\mu$ into the two parts as
\begin{gather}
	\boldsymbol{L}_\mu = \boldsymbol{L}_\mu^\parallel + \boldsymbol{L}_\mu^\perp
	,
\end{gather}
where
\begin{gather}
	\boldsymbol{L}_\mu^\parallel \equiv  (\boldsymbol{n} \cdot \boldsymbol{L}_\mu) \boldsymbol{n}
	\label{L_mu_para}
\end{gather}
is the part parallel to $\boldsymbol{n}$ and
\begin{gather}
	\boldsymbol{L}_\mu^\perp \equiv \boldsymbol{L}_\mu - \boldsymbol{L}_\mu^\parallel
	\label{L_mu_perp}
\end{gather}
is that perpendicular to $\boldsymbol{n}$.

The two-component spinor for the spin-up and spin-down states for an arbitrary quantization axis $\boldsymbol{n}$ are given by
\begin{gather}
	| \boldsymbol{n} \rangle =
	\begin{pmatrix}
		\cos ( \theta/2 ) \\
		e^{i \phi} \sin ( \theta/2 ) \\
	\end{pmatrix}
	, \,
	| -\boldsymbol{n} \rangle =
	\begin{pmatrix}
		\sin ( \theta/2 ) \\
		-e^{i \phi} \cos ( \theta/2 ) \\
	\end{pmatrix}
	,
	\label{two-comp_spinor}
\end{gather}
where $\theta$ and $\phi$ are the polar and the azimuthal angles of $\boldsymbol{n}$, respectively.
It is easily confirmed that $\boldsymbol{n} \cdot \langle \pm \boldsymbol{n} | \boldsymbol{S} | \mp \boldsymbol{n} \rangle = 0$
and we obtain the relation
\begin{gather}
	\boldsymbol{n} \cdot \boldsymbol{S} = \frac{ P_\uparrow - P_\downarrow }{2}
	.
\end{gather}
Using this relation,
the contribution from the parallel component for the energy correction is calculated from eqs. (\ref{def_Q_SOI}) and (\ref{L_mu_para}) as,
\begin{gather}
	\langle Q_\mu   \boldsymbol{L}_\mu^\parallel \cdot \boldsymbol{S} Q_\mu \rangle
		=
		\frac{1}{2}
		\boldsymbol{n} \cdot ( \xi_\mu^\uparrow \langle \boldsymbol{L}_\mu^\uparrow \rangle - \xi_\mu^\downarrow \langle \boldsymbol{L}_\mu^\downarrow \rangle )
	,
	\label{correction_spin_conserving}
\end{gather}
where $\langle \boldsymbol{L}_\mu^\sigma \rangle \equiv \langle \boldsymbol{L}_\mu P_\sigma \rangle$ ($\sigma = +, -$)
is the OAM acquired via the perturbation by the electrons of spin $\sigma$. 
$\langle \cdot \rangle$ represents the expectation value with respect to the perturbed ground state.
It can also be confirmed for an arbitrary $\sigma$ that 
$\langle P_\sigma Q_\mu   \boldsymbol{L}_\mu^\parallel \cdot \boldsymbol{S} Q_\mu P_{-\sigma} \rangle = 0$,
which means that the contribution from the parallel component within the second-order perturbation contains only the spin-conserving transitions.

Using the relation
\begin{gather}
	P_\sigma \boldsymbol{S} P_\sigma = \frac{\sigma}{2} \boldsymbol{n} P_\sigma
	,
	\label{PSP_snP}
\end{gather}
the contribution from the perpendicular component for the energy correction is calculated from eqs. (\ref{def_Q_SOI}) and (\ref{L_mu_perp}) as,
\begin{gather}
	\langle Q_\mu   \boldsymbol{L}_\mu^\perp \cdot \boldsymbol{S} Q_\mu \rangle
		=
		\sqrt{ \xi_\mu^\uparrow \xi_\mu^\downarrow }
		\langle
			\boldsymbol{L}_\mu \cdot \boldsymbol{T}
		\rangle
	,
	\label{correction_spin_flip}
\end{gather}
where we have defined the hermitian operator
\begin{gather}
	\boldsymbol{T} (\boldsymbol{n} )
		\equiv
		P_\uparrow \boldsymbol{S} P_\downarrow + P_\downarrow \boldsymbol{S} P_\uparrow
	.
	\label{def_T}
\end{gather}
It can also be confirmed for an arbitrary $\sigma$ that 
$\langle P_\sigma Q_\mu   \boldsymbol{L}_\mu^\perp \cdot \boldsymbol{S} Q_\mu P_{\sigma} \rangle = 0$,
which means that the contribution from the perpendicular component within the second-order perturbation contains only the spin-flip transitions.

By substituting eqs. (\ref{correction_spin_conserving}) and (\ref{correction_spin_flip}) into eq. (\ref{delta_E_0}),
we obtain the correction to the energy of the ground state,
\begin{gather}
	\delta E_0 (\boldsymbol{n})
		=
		\frac{1}{4} \sum_\mu
		\boldsymbol{n} \cdot ( \xi_\mu^\uparrow \langle \boldsymbol{L}_\mu^\uparrow \rangle - \xi_\mu^\downarrow \langle \boldsymbol{L}_\mu^\downarrow \rangle )
	\nonumber \\
		+ \frac{1}{2} \sum_\mu
			\sqrt{ \xi_\mu^\uparrow \xi_\mu^\downarrow }
			\langle \boldsymbol{L}_\mu \cdot \boldsymbol{T} \rangle 
	\label{corr_formula}
	.
\end{gather}
This expression is exact within the second-order perturbation theory.
It is clear that $\delta E_0 (\boldsymbol{n})$ consists of the three kinds of contributions:
The spin-conserving two transitions of the spin-up electrons,
those of the spin-down electrons,
and the spin-flip two transitions of the electrons of both spin directions.
When $\boldsymbol{n} = \boldsymbol{e}_z$, for example,
the spin-conserving contributions in eq. (\ref{corr_formula}) symbolically correspond to the quantity
$(\xi / 2) \langle L_z S_z \rangle$,
while the spin-flip contributions to
$( \xi / 2 ) \langle L_x S_x + L_y S_y \rangle$.
For an exchange splitting $\Delta E_\mathrm{ex}$,
the spin-flip contribution in eq. (\ref{corr_formula}) is on the order of $( \xi/\Delta E_\mathrm{ex})^2$, expected to be much smaller than the spin-conserving contributions.
If it is true, we could neglect the spin-flip contribution.
Furthermore,
when the majority spin bands, assumed to be spin-up here, are completely filled and the exchange splitting are very large,
the net OAM of the perturbed spin-up states vanishes.
In such a case, the formula eq. (\ref{corr_formula}) reads
\begin{gather}
	\delta E_0 (\boldsymbol{n}) \approx
		-\frac{1}{4} \sum_\mu \xi_\mu \boldsymbol{n} \cdot \langle \boldsymbol{L}_\mu \rangle
	,
	\label{Bruno_formula}
\end{gather}
which is nothing but the well known Bruno's formula\cite{bib:2143}.

Since the spin wave function of each electron is $| \boldsymbol{n} \rangle$ or $| -\boldsymbol{n} \rangle$ in the unperturbed system,
$\langle \boldsymbol{T} \rangle = \mathcal{O} (\xi_\mu)$.
Eqs (\ref{def_Q_SOI}) and (\ref{PSP_snP}) thus lead to
\begin{gather}
	\langle Q_\mu  \boldsymbol{S} Q_\mu \rangle
		=
		\frac{1}{2}
		\langle
			\xi_\mu^\uparrow P_\uparrow - \xi_\mu^\downarrow P_\downarrow
		\rangle
		\boldsymbol{n}
		+ \mathcal{O} (\xi_\mu^2)
	.
	\label{ex_QSQ}
\end{gather}
Remembering that 
$\langle \boldsymbol{L}_\mu^\perp \rangle = \mathcal{O} (\xi_\mu)$
since the OAM in the unperturbed system vanishes,
we obtain
$\langle \boldsymbol{L}_\mu^\perp \rangle \cdot \langle Q_\mu  \boldsymbol{S} Q_\mu \rangle = \mathcal{O} (\xi_\mu^3)$.
With this relation, the contribution to the energy correction from the perpendicular component is rewritten as
\begin{gather}
	\langle Q_\mu   \boldsymbol{L}_\mu^\perp \cdot \boldsymbol{S} Q_\mu \rangle
		=
		\langle Q_\mu  ( \boldsymbol{L}_\mu^\perp - \langle \boldsymbol{L}_\mu^\perp \rangle ) \cdot \boldsymbol{S} Q_\mu \rangle
\end{gather}
within the second-order perturbation.
Since the operator $\boldsymbol{L}_\mu^\perp$ appears in this expression as the difference between itself and its expectation value,
the contribution from the perpendicular component comes only from the quantum fluctuation effect.
This contribution does not vanish in general even when $\langle \boldsymbol{L}_\mu^\perp \rangle$ vanishes.
This result means that the contribution from the spin-flip transitions to the MCA in a ferromagnet is purely of quantum nature.
On the other hand, it is easily confirmed that 
$\langle \boldsymbol{L}_\mu^\parallel \rangle \cdot \langle Q_\mu   \boldsymbol{S} Q_\mu \rangle  = \mathcal{O} (\xi_\mu^2)$,
indicating that the mean-field effect can be present in the contribution from the spin-conserving transitions.

\subsection{OAM of single-particle states}

In the derivation of the second-order perturbation formula above,
the system was assumed to be described by a single many-body wave function.
In solid state physics, however,
the single-particle picture is often employed for a periodic system,
in which the system consists of the single-particle states whose occupation numbers are determined according to the Fermi level.
The net OAM of the system in such a case is calculated as the sum of the contributions from the occupied single-particle states.
The formula derived above does not take into account the variation in the Fermi level via the perturbation.
Here we examine the behavior of the net OAM of a periodic system with SOI in detail.

\subsubsection{Major contributions to net OAM}

We assign each of the perturbed single-particle states to four groups according to its occupation before and after the SOI is turned on as follows.
We denote an occupied perturbed state
by $| \psi_i^{\mathrm{occ (unocc)} \to \mathrm{occ}} \rangle$ if it was an unperturbed occupied (unoccupied) one.
We denote an unoccupied perturbed state by $| \psi_i^{\mathrm{occ (unocc)} \to \mathrm{unocc}} \rangle$ similarly.
We write the OAM of the $i$th single-particle state as the sum of the contributions of all orders in the SOI as
$\langle \psi_i^{a \to b} | \boldsymbol{L} | \psi_i^{a \to b} \rangle \equiv \boldsymbol{L}_i^{a \to b} \equiv \sum_{n = 0}^\infty \boldsymbol{L}_{i (n) }^{a \to b}$ ($a, b = \mathrm{occ, unocc}$).
The net OAM of the perturbed system is then written as
$\langle \boldsymbol{L} \rangle =  \sum_i \boldsymbol{L}_i^{\mathrm{unocc} \to \mathrm{occ}}  + \sum_i \boldsymbol{L}_i^{\mathrm{occ} \to \mathrm{occ}}$,
while the quench of OAM in the unperturbed system is expressed as
$0 = \sum_i \boldsymbol{L}_{i (0)}^{\mathrm{occ} \to \mathrm{occ}} + \sum_i \boldsymbol{L}_{i (0)}^{\mathrm{occ} \to \mathrm{unocc}} $.
We can thus write
$\langle \boldsymbol{L} \rangle = \sum_i \boldsymbol{L}_i^{\mathrm{unocc} \to \mathrm{occ}}  - \sum_i \boldsymbol{L}_{i (0)}^{\mathrm{occ} \to \mathrm{unocc}} + \sum_{i, n \ne 0} \boldsymbol{L}_{i (n)}^{\mathrm{occ} \to \mathrm{occ}}$.
Extracting the lowest-order contributions from each term on the right hand side of this expression,
we can write the net OAM as
\begin{gather}
	\langle \boldsymbol{L} \rangle
	\approx
		\sum_i \boldsymbol{L}_{i (0)}^{\mathrm{unocc} \to \mathrm{occ}}
		-\sum_i \boldsymbol{L}_{i (0)}^{\mathrm{occ} \to \mathrm{unocc}}
		+ \sum_i \boldsymbol{L}_{i (1)}^{\mathrm{occ} \to \mathrm{occ}}
	.
	\label{net_OAM_3_terms}
\end{gather}
The first (second) summation on the right hand side of this expression
is roughly proportional to the number of occupied (unoccupied) perturbed states which was unoccupied (occupied) unperturbed states.
These contributions are determined not only by the perturbed energy eigenvalues of the single-particle states,
but also by the perturbed Fermi level, which is determined by the perturbed energy eigenvalues.
The order of perturbation contributing to the Fermi level and the net OAM is thus not trivial
even when the correction to the energy eigenvalues is of the first order.

Since each of the unperturbed single-particle states does not contain the zeroth order contribution for the expectation value of the operator $\boldsymbol{L} \cdot \boldsymbol{T}$,
the leading contributions to the operator come only from the states which are occupied before and after the SOI is turned on:
\begin{gather}
	\langle \boldsymbol{L} \cdot \boldsymbol{T} \rangle
	\approx
		\sum_i ( \boldsymbol{L} \cdot \boldsymbol{T} )_{i (1)}^{\mathrm{occ} \to \mathrm{occ}}
	,
	\label{net_LT_1_term}
\end{gather}
to be compared with the expression for the net OAM, eq. (\ref{net_OAM_3_terms}).

\subsubsection{Parabolic-bands model}

\begin{figure}[htbp]
\begin{center}
\includegraphics[keepaspectratio,width=7cm]{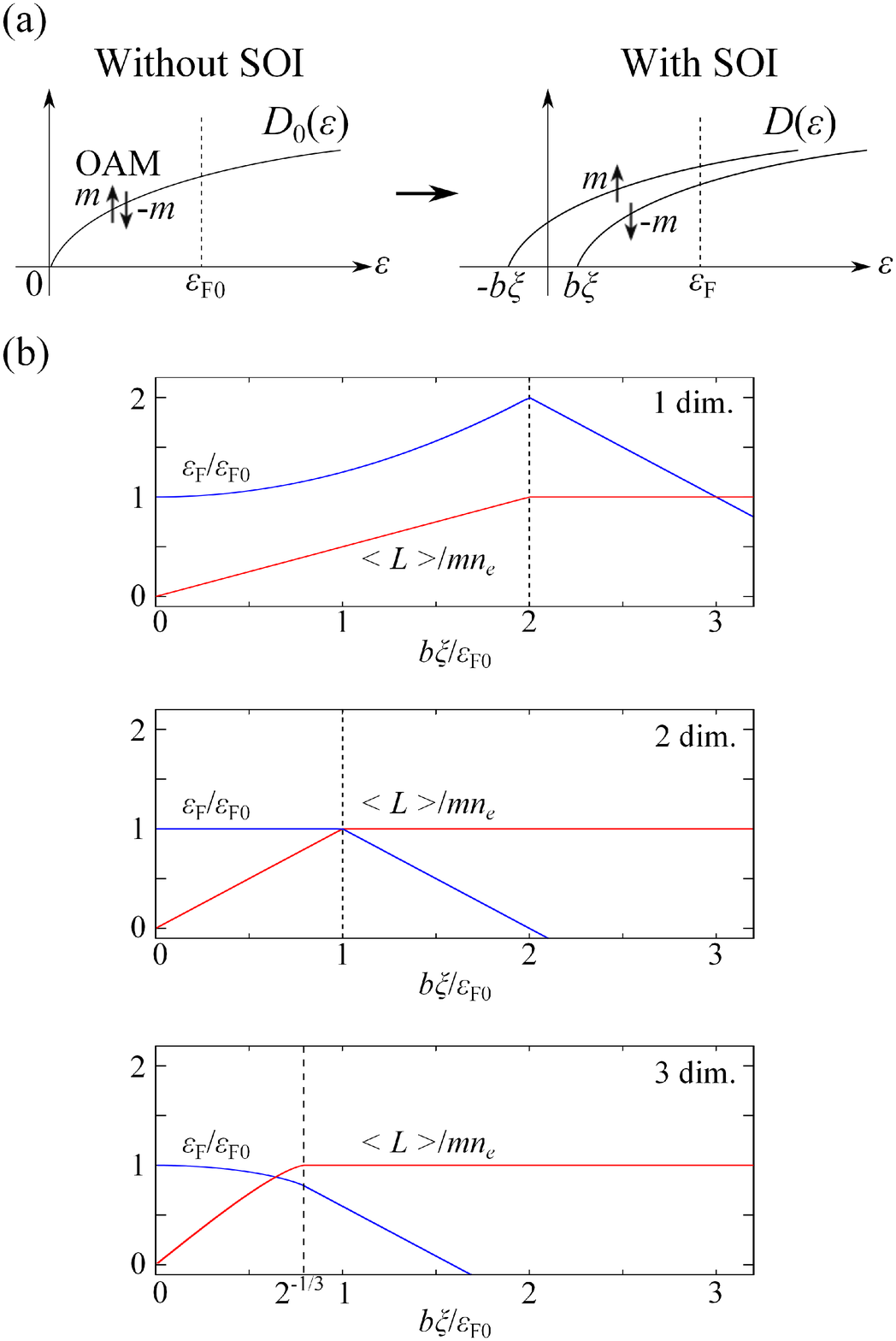}
\end{center}
\caption{
(a)
(Color online)
Schematic illustration of the density of states for the model consisting of two parabolic bands.
The origin of energy is set to the common bottom of the unperturbed bands.
Each state in one of the bands has an OAM $m$,
while each state in the other band has an OAM $-m$.
The unperturbed two bands coincide with each other and thus the net OAM vanishes.
With the SOI turned on,
the band with the OAM $m$ undergoes a rigid shift $-b \xi$ and the other band with $-m$ underdoes a rigid shift $b \xi$.
The unperturbed Fermi level $\varepsilon_{\mathrm{F} 0}$ is changed to the perturbed one $\varepsilon_\mathrm{F}$
for the conservation of the number of electrons.
(b)
The Fermi level and the net OAM of the perturbed systems for one-, two-, and three-dimensional systems within our model
as functions of the relative strength $b \xi / \varepsilon_{\mathrm{F} 0}$ of SOI with respect to the unperturbed Fermi level.
For each dimension, dashed vertical line represents the critical value of the SOI strength above which the band having OAM $-m$ is empty.
}
\label{fig_dos_oam}
\end{figure}

To see the behavior of the net OAM acquired by a periodic system via the change in its Fermi level,
here we examine a model consisting of two parabolic bands whose bottoms are close to the Fermi level.
We assume that the two bands without SOI coincide with each other.
We further assume that each unperturbed single-particle state in one band has an intrinsic OAM $m$
and that in the other band has an OAM with the same magnitude but in the opposite direction, $-m$.
Such a situation may not be very special since the net OAM of a generic system without SOI vanishes, as stated above.
For simplicity, we consider a case in which the band with the OAM $m$ undergoes a rigid shift $-b \xi$ as the perturbation, first-order in the strength $\xi$ of the SOI,
while the other band with $-m$ underdoes a rigid shift $b \xi$.
$b$ is a dimensionless positive constant.
The schematic illustration of the model is shown in Fig. \ref{fig_dos_oam} (a).
Since the expression of DOS for a parabolic band is known\cite{bib:Grosso_Parravicini},
the exact expressions of the Fermi level and the net OAM as functions of the SOI strength $\xi$ can be derived.
They are provided in Appendix.
While the leading order of variation in the Fermi level as $\xi$ is changed
depends on the dimensionality of the system [see eqs. (\ref{ef_para_1dim_small_xi}), (\ref{ef_para_2dim_small_xi}), and (\ref{ef_para_3dim_small_xi_approx})],
interestingly,
that in the net OAM is of the first order in $\xi$ regardless of the dimensionality [see eqs. (\ref{oam_para_1dim_small_xi}), (\ref{oam_para_2dim_small_xi}), and (\ref{oam_para_3dim_small_xi_approx})].
The Fermi level and the net OAM of the perturbed systems for one-, two-, and three-dimensional systems within our model
are plotted in Fig. \ref{fig_dos_oam} (b) as functions of the relative strength $b \xi / \varepsilon_{\mathrm{F} 0}$ of SOI with respect to the unperturbed Fermi level $\varepsilon_{\mathrm{F} 0}$ measured from the common bottom of the unperturbed bands.
For a fixed value of the number $n_e$ of electrons,
there exists the critical strength of SOI above which the band having OAM $-m$ is empty and thus the net OAM is saturated.
It is found that, even when the strength $\xi$ of SOI itself is large,
the net OAM is small if the unperturbed Fermi level is much larger than $\xi$.
Similar discussion is also applicable to the top of parabolic bands by considering the number of holes.

If we believe that the mechanism of the appearance of net OAM for parabolic bands examined above is true at least qualitatively also for systems having generic bands,
we understand that the states which are close not only to the Fermi level but also to the bottoms or the tops of bands can contribute to the appearance of the net OAM.
Such contributions are assigned to the first and the second terms in eq. (\ref{net_OAM_3_terms})
and they give the first-order contributions in SOI, as demonstrated above.

\subsection{Connection between Perturbation Formula and DFT calculations}

In the present study, we perform self-consistent FR DFT calculations.
The resultant two-component Bloch states, which are the energy eigenstates of the FR Kohn-Sham Hamiltonian,
are in general the mixture of the spin-up and -down states with respect to a given quantization axis $\boldsymbol{n}$.
To evaluate the right hand side of the second-order perturbation formula, eq. (\ref{corr_formula}),
we define the OAM matrix of the atom $\mu$ using the occupied FR eigenstates as
\begin{gather}
	\boldsymbol{\mathcal{L}}_{\mu}^{ \tau \tau'}
		\equiv \sum_{i \in \mathrm{occ.}} \langle \psi_{i \tau'} | \boldsymbol{L}_\mu  | \psi_{i \tau} \rangle
	\label{def_OAMmat}
\end{gather}
for $\tau, \tau' = \alpha, \beta$.
The spin indices $\alpha$ and $\beta$ used in a DFT calculation does not necessarily correspond to the eigenstates of spin directions for the quantization axis $\boldsymbol{n}$.
$\boldsymbol{\mathcal{L}}_{\mu}$ is a hermitian matrix with respect to the indices $\tau$ and $\tau'$.
By using the matrix representations of the spin projection operator $P_\sigma$ ($\sigma = \uparrow, \downarrow$)
and the spin-flip operator $\boldsymbol{T}$,
we obtain
\begin{gather}
	\langle \boldsymbol{L}_\mu^\sigma \rangle
		= \mathrm{Tr} ( P_{\sigma} \boldsymbol{\mathcal{L}}_{ \mu} )
	,
	\\
	\langle \boldsymbol{L}_\mu \cdot \boldsymbol{T} \rangle
		= \mathrm{Tr} ( \boldsymbol{\mathcal{L}}_{ \mu} \cdot \boldsymbol{T} )
	.
\end{gather}
Using the two-component spinor, eq (\ref{two-comp_spinor}),
the matrix representation of $P_\sigma$'s are written as
\begin{gather}
	P_\uparrow 
	=
	\begin{pmatrix}
		\cos^2 \frac{\theta}{2} & e^{-i \phi} \cos \frac{\theta}{2} \sin \frac{\theta}{2} \\
		e^{i \phi} \cos \frac{\theta}{2} \sin \frac{\theta}{2} & \sin^2 \frac{\theta}{2} \\
	\end{pmatrix}
	, \, \\
	P_\downarrow
	=
	\begin{pmatrix}
		\sin^2 \frac{\theta}{2} & -e^{-i \phi} \cos \frac{\theta}{2} \sin \frac{\theta}{2} \\
		-e^{i \phi} \cos \frac{\theta}{2} \sin \frac{\theta}{2} & \cos^2 \frac{\theta}{2} \\
	\end{pmatrix}
	.
\end{gather}
The matrix representation of $\boldsymbol{T}$, defined as eq. (\ref{def_T}), are written as
\begin{gather}
	T_x = \frac{1}{2}
	\begin{pmatrix}
		-\frac{1}{2} \sin 2 \theta \cos \phi & 1 - \frac{1}{2} \sin^2 \theta (1 + e^{-2 i \phi}) \\
		1 - \frac{1}{2} \sin^2 \theta (1 + e^{2 i \phi}) & \frac{1}{2} \sin 2 \theta \cos \phi \\
	\end{pmatrix}
	\\
	T_y = \frac{1}{2}
	\begin{pmatrix}
		-\frac{1}{2} \sin 2 \theta \sin \phi & -i + \frac{i}{2} \sin^2 \theta (1 - e^{-2 i \phi}) \\
		i - \frac{i}{2} \sin^2 \theta (1 - e^{2 i \phi}) & \frac{1}{2} \sin 2 \theta \sin \phi \\
	\end{pmatrix}
	\\
	T_z = \frac{1}{2}
	\begin{pmatrix}
		\sin^2 \theta & -\frac{1}{2} \sin 2 \theta e^{-i \phi} \\
		-\frac{1}{2} \sin 2 \theta e^{i \phi} & -\sin^2 \theta  \\
	\end{pmatrix}
\end{gather}
The explicit expressions of the OAM induced by the perturbation to the spin-up and -down wave functions are
\begin{gather}
	\langle \boldsymbol{L}_\mu^\uparrow \rangle
		= \boldsymbol{\mathcal{L}}_\mu^{\alpha \alpha} \cos^2 \frac{\theta}{2}
		+ \boldsymbol{\mathcal{L}}_\mu^{\beta \beta} \sin^2 \frac{\theta}{2}
		+ \mathrm{Re} ( \boldsymbol{\mathcal{L}}_\mu^{\alpha \beta} e^{-i \phi} ) \sin \theta
	, \\
	\langle \boldsymbol{L}_\mu^\downarrow \rangle
		= \boldsymbol{\mathcal{L}}_\mu^{\alpha \alpha} \sin^2 \frac{\theta}{2}
		+ \boldsymbol{\mathcal{L}}_\mu^{\beta \beta} \cos^2 \frac{\theta}{2}
		- \mathrm{Re} ( \boldsymbol{\mathcal{L}}_\mu^{\alpha \beta} e^{-i \phi} ) \sin \theta
	,
\end{gather}
and hence $\langle \boldsymbol{L}_\mu \rangle = \langle \boldsymbol{L}_\mu^\uparrow \rangle + \langle \boldsymbol{L}_\mu^\downarrow \rangle = \mathrm{Tr} \boldsymbol{\mathcal{L}}_\mu$.

The explicit expressions for the evaluation of OAM matrices in a DFT calculation using a plane-wave basis set are provided in Appendix.

\section{Applications}

\subsection{Tight-binding calculation for a $d$ orbital chain}

As the first example, we examine the electronic structure of a $d$ orbital chain by performing tight-binding calculations.
Each site on the chain is distant from its neighbor by $a$ in the $z$ direction.

\subsubsection{Hamiltonian and electronic band structure}

\begin{figure}[htbp]
\begin{center}
\includegraphics[keepaspectratio,width=5cm]{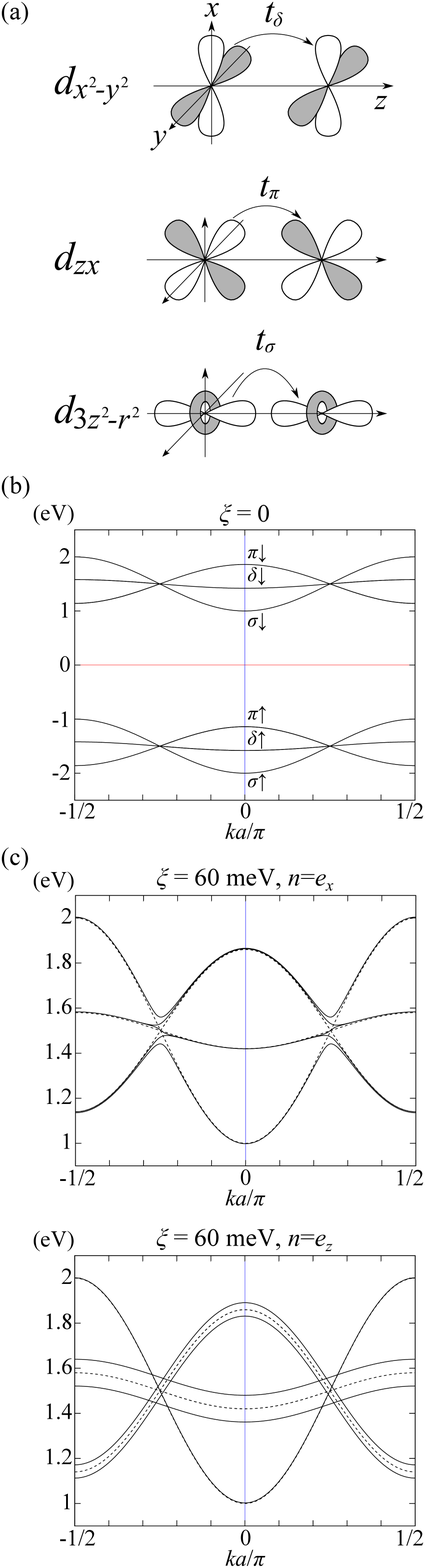}
\end{center}
\caption{
(a)
Schematic illustration of the transfer integrals used for the $d$ orbital chain.
(b)
The nonrelativistic ($\xi = 0$) electronic band structure of the chain accommodating six electrons per site.
The transfer integrals used are $t_\delta = -0.04, t_\pi = 0.18$, and $t_\sigma = -0.25$ eV.
The exchange splitting is set to $\Delta_\mathrm{ex} = 3$ eV.
(c)
Solid curves represent the relativistic electronic band structures with $\xi = 60$ meV for $\boldsymbol{n} = \boldsymbol{e}_x$ and $\boldsymbol{e}_z$ on the upper and the lower panels, respectively.
The relativistic bands coming from the nonrelativistic spin-down states, whose direction is $-\boldsymbol{n}$, are shown in the figures.
The nonrelativistic bands are also shown as the dashed curves.
}
\label{fig_chain_bands}
\end{figure}

Only the transfer integrals between neighboring sites are considered here.
We set their values as
$t_\delta = -0.04, t_\pi = 0.18$, and $t_\sigma = -0.25$ eV [see Fig. \ref{fig_chain_bands} (a)].
The exchange splitting is set to $\Delta_\mathrm{ex} = 3$ eV.
These values are the same as in the tight-binding analysis done by Wang \etal\cite{bib:1976} for a diatomic molecule of iron.

We denote the $d_i$ orbital ($i = xy, yz, zx, x^2 - y^2, 3z^2 - r^2$) with its spin direction $\boldsymbol{n}$ ($-\boldsymbol{n}$) by $d_i^\uparrow$ ($d_i^\downarrow$).
We arrange the Bloch sums of these orbitals on the chain as $\{ d_{xy}^\uparrow, d_{yz}^\uparrow, d_{zx}^\uparrow, d_{x^2 - y^2}^\uparrow, d_{3 z^2 - r^2}^\uparrow,  d_{xy}^\downarrow, d_{yz}^\downarrow, d_{zx}^\downarrow, d_{x^2 - y^2}^\downarrow, d_{3 z^2 - r^2}^\downarrow \}$.
With this basis set, the tight-binding Hamiltonian matrix for a wave number $k$ in the one-dimensional Brillouin zone is written as
\begin{gather}
	H(k) =
	\begin{pmatrix}
		V (k) - \frac{\Delta_{\mathrm{ex}}}{2} + H_\mathrm{SO}^{\uparrow \uparrow}  & H_\mathrm{SO}^{\uparrow \downarrow}  \\
		H_\mathrm{SO}^{\downarrow \uparrow}  & V (k) + \frac{\Delta_{\mathrm{ex}}}{2} + H_\mathrm{SO}^{\downarrow \downarrow}  \\
	\end{pmatrix}
	,
	\label{chain_H_mat}
\end{gather}
where
\begin{gather}
	V(k) = 2 \cos k a
	\begin{pmatrix}
		t_\delta & & & & \\
		& t_\pi & & & \\
		& & t_\pi & & \\
		& & & t_\delta & \\
		& & & & t_\sigma 
	\end{pmatrix}
	.
\end{gather}
Each component on the right hand side of eq. (\ref{chain_H_mat}) is a $5 \times 5$ matrix.
The transfer integral is nonzero only between the neighboring same orbitals due to the symmetry.
$H_\mathrm{SO}^{\sigma \sigma'}$ ($\sigma, \sigma' = \uparrow, \downarrow$) represents the matrix elements of the SOI Hamiltonian eq. (\ref{def_H_SOI}) with the strength $\xi$ common to both directions of spin.
The electronic band structure of the chain is obtained by numerically diagonalizing $H(k)$ for each $k$.

The nonrelativistic ($\xi = 0$) electronic band structure is shown in Fig. \ref{fig_chain_bands} (b).
The ten bands consist of two band groups corresponding to the spin-up and -down states, each of which contains five bands.
The five bands in each band group are made up of two $\delta$, two $\pi$, and one $\sigma$ bands due to the axial symmetry of the chain.

The relativistic electronic band structures with $\xi = 60$ meV are shown in Fig. \ref{fig_chain_bands} (c) for $\boldsymbol{n} = \boldsymbol{e}_x$ and $\boldsymbol{e}_z$.
It is seen that the features of band structures for the different spin directions are different from each other due to the SOI.

\subsubsection{OAM}

\begin{figure*}[htbp]
\begin{center}
\includegraphics[keepaspectratio,width=17.5cm]{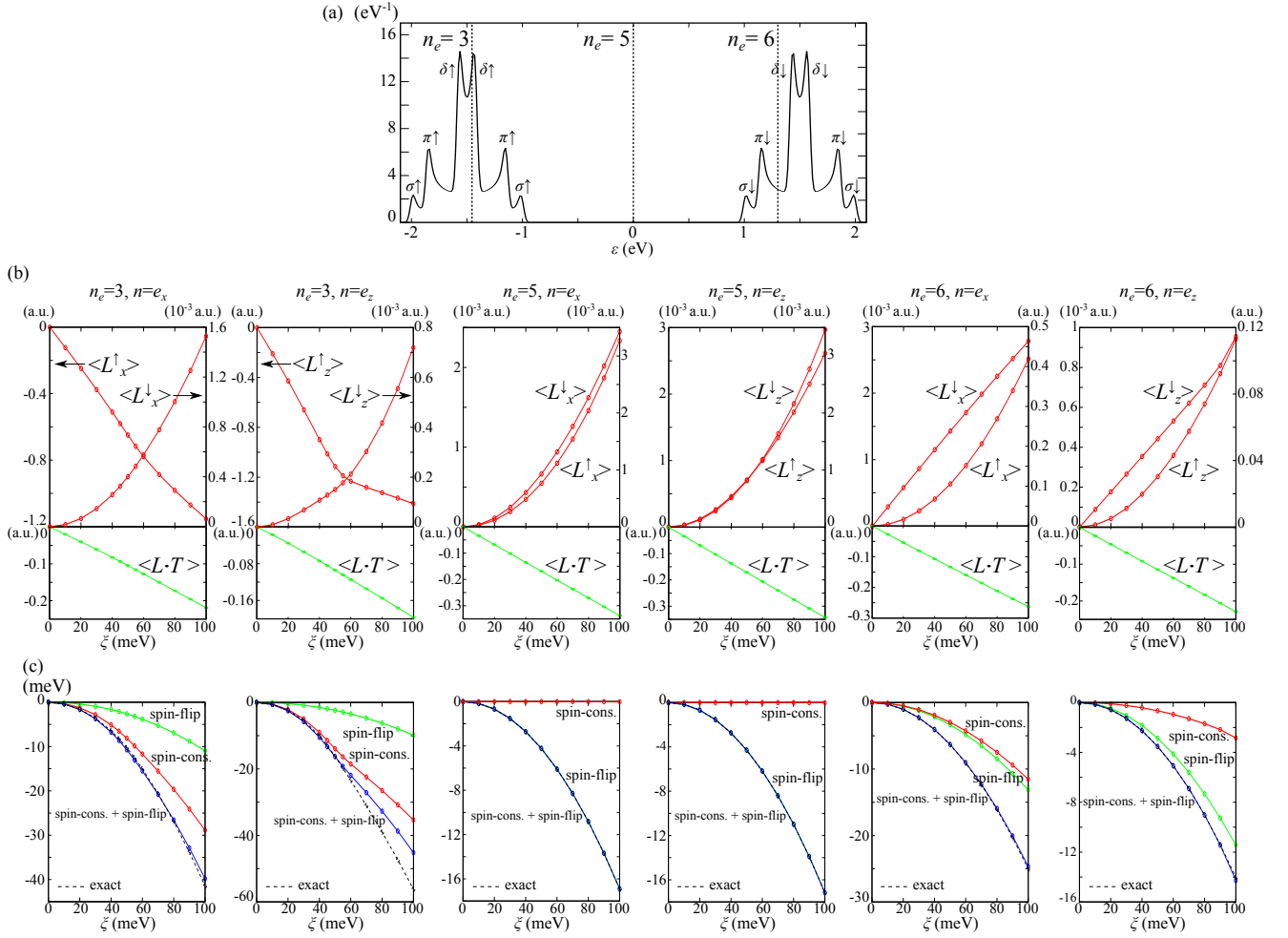}
\end{center}
\caption{
(a)
The density of states for the nonrelativistic band structure of the chain.
Gaussian broadening was used.
The vertical dashed lines represent the Fermi levels for the numbers of electrons per site  $n_e = 3, 5$,and $6$.
For $n_e = 5$, the Fermi level lies in the exchange splitting.
(b)
The net OAM and the expectation values of $\langle \boldsymbol{L} \cdot \boldsymbol{T} \rangle$ as functions of $\xi$ for the spin directions along the $x$ and the $z$ axes.
(c) The contributions to the second-order perturbation formula as functions of $\xi$, together with the numerically exact correction to the total energy.
}
\label{fig_chain_OAM}
\end{figure*}

Here we examine the behavior of the net OAM acquired by the system via the variation in the strength $\xi$ of SOI and the Fermi level.
The density of states for the nonrelativistic band structure of the chain is plotted in Fig. \ref{fig_chain_OAM} (a).
For the numbers of electrons per site $n_e = 3, 5$, and $6$,
the net OAM and the expectation values of $\langle \boldsymbol{L} \cdot \boldsymbol{T} \rangle$ as functions of $\xi$ are plotted in Fig. \ref{fig_chain_OAM} (b).

Let us analyze the OAM in detail for the case of the electron spins in the $z$ direction ($\boldsymbol{n} = \boldsymbol{e}_z$)
since the exact eigenstates of the Hamiltonian in this case can be obtained analytically, as provided in Appendix.
The lowest-order contribution to the expectation value of $L_z$ of every energy eigenstate in this case is of the second order in $\xi$ [see eqs. from (\ref{chain_z_Lz_2nd_start}) to (\ref{chain_z_Lz_2nd_end})].
Despite that,
the behaviors of $\langle L^\uparrow_z \rangle$ for $n_e = 3$ and $\langle L^\downarrow_z \rangle$ for $n_e = 6$ are obviously not quadratic, as seen in Fig. \ref{fig_chain_OAM} (b).
Those of $\langle L^\uparrow_z \rangle$ and $\langle L^\downarrow_z \rangle$ for $n_e = 5$ are, on the other hand, quadratic.
These results clearly indicate that 
the major contributions for the $n_e=3$ and $6$ cases are from
the perturbed states that has moved through the Fermi level when the SOI was turned on.

We define the OAM density as
\begin{gather}
	D_{\boldsymbol{L}}  (\varepsilon )
	\equiv
		\sum_{k, i} \langle \psi_{k i} | \boldsymbol{L} | \psi_{k i} \rangle \delta (\varepsilon - \varepsilon_{ki})
		f (\varepsilon ; \varepsilon_\mathrm{F} )
	,
\end{gather}
calculated from the perturbed single-particle states and the corresponding enregy eigenvalues for a given quantization axis $\boldsymbol{n}$.
$f (\varepsilon ; \varepsilon_\mathrm{F} )$ is the Fermi distribution function with the Fermi level $\varepsilon_\mathrm{F}$ of the perturbed system.
In addition, we define the accumulated OAM density as
\begin{gather}
	\boldsymbol{L}_\mathrm{acc} (\varepsilon ) 
	\equiv
	\int_{-\infty}^\varepsilon  d \varepsilon'
		D_{\boldsymbol{L}}  (\varepsilon' )
	.
\end{gather}
The net OAM is clearly the accumulated OAM density up to the Fermi level: $\langle \boldsymbol{L} \rangle = \boldsymbol{L}_\mathrm{acc} (\varepsilon_\mathrm{F} )$.
For the operator $\boldsymbol{L} \cdot \boldsymbol{T}$,
we define $D_{\boldsymbol{L} \cdot \boldsymbol{T} } (\varepsilon )$ and  $\boldsymbol{L} \cdot \boldsymbol{T}_\mathrm{acc} (\varepsilon )$  similarly to the OAM.

\begin{figure*}[htbp]
\begin{center}
\includegraphics[keepaspectratio,width=17cm]{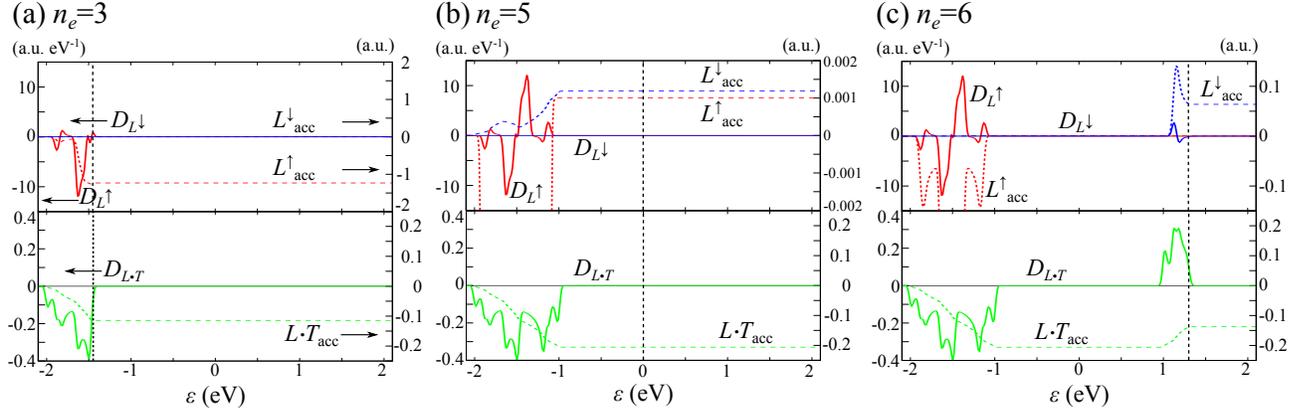}
\end{center}
\caption{
(Color online)
The OAM densities (solid curves) and their accumulated values (dashed curves) in the perturbed system with $\boldsymbol{n} = \boldsymbol{e}_z$ and $\xi = 60$ meV as functions of energy for $n_e = 3, 5$, and $6$.
The densities and the accumulated values of the operator $\boldsymbol{L} \cdot \boldsymbol{T}$ are also shown.
For each of the $n_e$'s, the vertical line represents the Fermi level of the unperturbed system.
}
\label{fig_chain_enres}
\end{figure*}

For $\boldsymbol{n} = \boldsymbol{e}_z$ and $\xi = 60$ meV, the OAM densities and their accumulated values as functions of energy for $n_e = 3, 5$, and $6$ are plotted in Fig. \ref{fig_chain_enres}.
The densities and the accumulated values of the operator $\boldsymbol{L} \cdot \boldsymbol{T}$ are also shown.
It is seen for $n_e = 3, 5$, and $6$ that $D_{L_z^\uparrow} (\varepsilon )$ oscillates strongly around the origin.
These large amplitudes come mainly from the intrinsic OAM of the single-particle states,
that is, the nonzero OAM present even when the SOI is absent [see eqs. from (\ref{chain_z_Lz_2nd_start}) to (\ref{chain_z_Lz_2nd_end})].

When the OAM density is integrated for $n_e = 5$, however, these zeroth contributions vanish.
Due to the absence of the first-order contributions [see eqs. from (\ref{chain_z_Lz_2nd_start}) to (\ref{chain_z_Lz_2nd_end})],
the leading contribution to the net OAM of the spin-up states, $\langle L_z^\uparrow \rangle$, for $n_e = 5$ is of second order in $\xi$,
leading to the rather small $\langle L_z^\uparrow \rangle$.
The net OAM of the spin-down states, $\langle L_z^\downarrow \rangle$, is also nonzero and of second order
since the lower five bands contain the spin-down components when the SOI is present.

For $n_e = 3$, $\langle L_z^\uparrow \rangle$ is much larger in magnitude than that in the case of $n_e = 5$.
It is because that $\langle L_z^\uparrow \rangle$ in this case comes mainly from the variation in the occupation numbers of the states near the Fermi level,
which corresponds to the first and second terms on the right hand side of eq. (\ref{net_OAM_3_terms}).
As stated above, those terms are of first-order in $\xi$.
Since each of the single-particle states does not contain the first-order contribution to $L_z$,
the magnitude of $L_{z \mathrm{acc}}^\uparrow (\varepsilon)$ becomes larger steeply as  $\varepsilon$ approaches the Fermi level, as seen in Fig. \ref{fig_chain_enres} (a).
$\langle L_z^\downarrow \rangle$ in this case is much smaller in magnitude than $\langle L_z^\uparrow \rangle$
since the spin-down components of the single-particle states in the lower five bands do not have the intrinsic OAM and thus $\langle L_z^\downarrow \rangle$ is of second order in $\xi$ as well as for $n_e = 5$.

For $n_e = 6$, $\langle L_z^\downarrow \rangle$ is much larger than $\langle L_z^\uparrow \rangle$ in contrast to the case of $n_e = 3$.
It is because that $\langle L_z^\downarrow \rangle$ in this case comes mainly from the variation in the occupation numbers of the states near the Fermi level.
$\langle L_z^\uparrow \rangle$ in this case is much smaller than $\langle L_z^\downarrow \rangle$
since the spin-up components of the single-particle states in the higher five bands do not have the intrinsic OAM and thus $\langle L_z^\uparrow \rangle$ is of second order in $\xi$ as well as for $n_e = 5$.
It is found that the magnitude of $\langle L_z^\downarrow \rangle$ for $n_e = 6$ is much smaller than that of $\langle L_z^\uparrow \rangle$ for $n_e = 3$.
We can understand this result by considering the parabolic-bands model discussed above,
which indicates that the Fermi level close to a band bottom leads to a larger net OAM.
As seen in Fig. \ref{fig_chain_OAM} (a), the Fermi level for $n_e = 3$ is closer to its nearest peak of DOS
than that for $n_e = 6$ is.

The leading contributions to $\langle \boldsymbol{L} \cdot \boldsymbol{T} \rangle$ from the energy eigenstates in this case are of first order in $\xi$ [see eqs. from (\ref{chain_z_LT_2nd_start}) to (\ref{chain_z_LT_2nd_end})].
Since the leading contributions to the net $\langle \boldsymbol{L} \cdot \boldsymbol{T} \rangle$ come from all the states below away the Fermi level [see eq. (\ref{net_LT_1_term})],
the variation in $\boldsymbol{L} \cdot \boldsymbol{T}_\mathrm{acc} (\varepsilon )$ with the increase in $\varepsilon$ is mild compared to that in $\boldsymbol{L}_\mathrm{acc} (\varepsilon )$ for all the $n_e$'s,
as seen in Fig. \ref{fig_chain_enres}.

\subsubsection{Analysis using the perturbation formula}

We define the total energy of the system simply as the sum of the energy eigenvalues of the occupied states.
The total-energy corrections due to the SOI calculated from the second-order perturbation formula, eq. (\ref{corr_formula}), are shown in Fig. \ref{fig_chain_OAM} (c) as functions of $\xi$.
It is seen for $n_e = 5$ that the values calculated using the second-order formula are in excellent agreement with the numerically exact values for $\xi$'s in the adopted range.
The agreement between the values obtained by the formula and the exact values for $n_e = 6$ are also rather good.
The accurate reproduction of the numerically exact values is achieved only when the spin-conserving and the spin-flip contributions are incorporated together,
which indicates that the the Bruno's formula, eq. (\ref{Bruno_formula}), does not capture the relativistic physics accurately in this case.

The deviation of the values obtained by the formula from the exact values is, however, found to be much larger for $n_e = 3$ than for $n_e = 6$.
These results can be understood via consideration similar to that of the net OAM.
For the case of $n_e = 5$, the Fermi level lies in the exchange gap,
which ensures that the SOI as perturbation does not allow the unperturbed states near the Fermi level to go through it when the SOI is turned on.
The occupation numbers of the unperturbed states are thus unchanged before and after the SOI is turned on,
so that the energy correction formula, eq. (\ref{corr_formula}),
immediately applies in such a case and gives the exact correction to the total energy within the second-order perturbation.
For the cases of $n_e = 3$ and $6$, on the other hand,
the Fermi level lies in the bands.
The effects of the variation in the occupation numbers of the states near the Fermi level via the SOI are thus present in such cases.
These effects are not taken into account in the formula at all, as discussed above.
The correction to the total energy of the system thus cannot be explained completely by the formula
even when the SOI is weak enough to be treated within second-order perturbation.

\subsection{Density Functional Theory Calculation for $L1_0$ Alloys}

As the second example, we perform first-principles electronic structure calculations based on DFT for five $L1_0$ alloys, FePt, CoPt, FePd, MnAl, and MnGa.
We examine their MCA systematically by employing the second-order perturbation formula.

\subsubsection{Crystal structure}

\begin{figure}[htbp]
\begin{center}
\includegraphics[keepaspectratio,width=6cm]{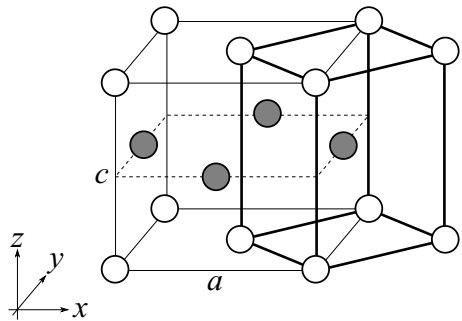}
\end{center}
\caption{
Crystal structure of an $L1_0$ alloy.
The white and the shaded balls represent atoms of different kinds.
The structure is regarded as stacked atomic layers spaced by $c$.
The thinner solid lines represent the conventional cell,
while the thicker solid lines represent the primitive cell.
The two kinds of atoms are located at the crystallographically equivalent positions.
The distance between the nearest neighbor atoms of the same kind is $a/\sqrt{2}$.
The $x$ and the $z$ directions correspond to the $[100]$ and the $[001]$ directions, respectively.
}
\label{fig_L10_lattice}
\end{figure}

The crystal structure of an $L1_0$ alloy is depicted in Fig. \ref{fig_L10_lattice}.
The basal lattice constants $a$ for the $L1_0$ alloys are fixed at the experimental values,
$3.8600$ \AA \, for FePt\cite{bib:2456},
$3.81$ \AA \, for CoPt\cite{bib:2480},
$3.89$ \AA \, for FePd\cite{bib:2674},
$3.92$ \AA \, for MnAl\cite{bib:2453}, and
$3.8974$ \AA \, for MnGa\cite{bib:2470},
throughout the present study.

\subsubsection{Computational details}

We adopt the projector augmented-wave (PAW) method\cite{bib:PAW} 
using the QMAS (Quantum MAterials Simulator) package\cite{bib:QMAS} 
within the local-spin-density approximation (LSDA).\cite{bib:LDA}  
We perform fully relativistic calculations for periodic systems\cite{bib:1504} using two-component pseudo Bloch wave functions as
\begin{gather}
	| \psi_{m \boldsymbol{k}} \rangle
	=
	\begin{pmatrix}
		| \psi_{m \boldsymbol{k} \alpha} \rangle \\
		| \psi_{m \boldsymbol{k} \beta} \rangle \\
	\end{pmatrix}
	,
\end{gather}
where $m$ and $\boldsymbol{k}$ are a band index and a wave vector, respectively.
$\alpha$ and $\beta$ are spin indices. 
In the present study, the pseudo wave functions are expanded in plane waves with an energy cutoff of 35 Ry.
The total energy of the system is calculated as a functional of the $2 \times 2$ density matrix defined in real space representation as
\begin{gather}
	\rho_{\tau \tau'} (\boldsymbol{r}) = \sum_{m, \boldsymbol{k}}^{\mathrm{occ.}}
		\psi_{m \boldsymbol{k} \tau} (\boldsymbol{r})^*
		\psi_{m \boldsymbol{k} \tau'} (\boldsymbol{r}),
\end{gather}
where $\tau, \tau' = \alpha, \beta$. 
In an FR calculation, noncollinear magnetism and spin-orbit interaction can be naturally introduced.

\begin{figure}[htbp]
\begin{center}
\includegraphics[keepaspectratio,width=5cm]{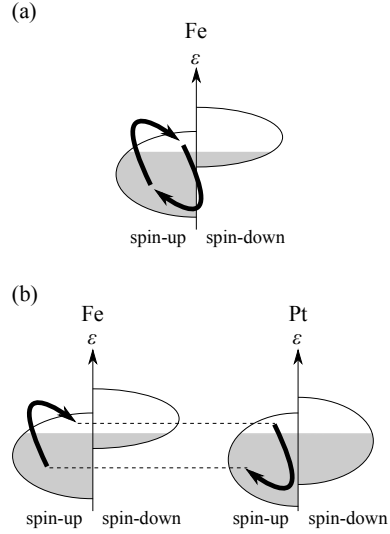}
\end{center}
\caption{
Schematic illustration of transitions in second-order perturbation processes for FePt.
The domes represent the density of states for each element,
whose shaded areas represent the occupied states.
For the spin-conserving transition of a spin-up electron in FePt,
two kinds of transitions exist.
(a)
The one occurs between the states only at the Fe atoms,
(b)
while the other occurs between the Fe and the Pt atoms.
Since the atomic orbitals of Fe and Pt are hybridized in the Bloch states,
an electron can travel between the atoms of different species via the second-order perturbation.
}
\label{fig_L10_transition}
\end{figure}

When we solve the Dirac equation for an isolated atom\cite{bib:Sakurai_Advanced} for the construction of a potential,
we can continuously move from the scalar relativistic ($\lambda = 0$) to the fully relativistic ($\lambda = 1$) equation
by varying the dimensionless parameter $\lambda$ of the SOI.\cite{bib:1920}
With turning on or off the SOI of each element for the potentials,
we can calculate the MCA energy only with the SOI around the atoms of a specific element.
For example, in FePt,
$E_\mathrm{MCA} (  \lambda_\mathrm{Fe} = 1, \lambda_\mathrm{Pt} = 0 )$ involves only the transitions between the states at the Fe atoms caused by the SOI around the Fe atoms.
We can extract the MCA energy coming only from the interspecies transitions as
\begin{gather}
	E_\mathrm{MCA} (\mathrm{between \, Fe \, and \, Pt})
		=
		E_\mathrm{MCA} ( \lambda_\mathrm{Fe} = 1, \lambda_\mathrm{Pt} = 1 )
	\nonumber \\
		- E_\mathrm{MCA} ( \lambda_\mathrm{Fe} = 1, \lambda_\mathrm{Pt} = 0 )
		- E_\mathrm{MCA} ( \lambda_\mathrm{Fe} = 0, \lambda_\mathrm{Pt} = 1 )
	.
	\label{MAE_interspecies}
\end{gather}
Although each transition in a perturbation process occurs at an atom due to the localized effectiveness of SOI, the interspecies contributions do not vanish in general.
It is because that the atomic orbitals of the different species in the $L1_0$ alloys extend to induce the hybridization with each other,
which allows an electron to travel between the atoms of different species, as illustrated in Fig. \ref{fig_L10_transition}.
It is clear that the interspecies contributions do not contain the influence of the first-order transitions.

The strengths $\xi$ of the SOI of each atom used for the second-order perturbation formula are estimated from the energy eigenvalues obtained in the DFT calculation for an isolated atom.
The low-energy expansion of the Dirac equation for a central potential $V$ leads to the following SOI Hamiltonian of the well known form\cite{bib:Sakurai_Advanced}:
\begin{gather}
	H_{\mathrm{SO}} = \frac{1}{2m^2 c^2} \frac{1}{r} \frac{\diff V}{\diff r} \boldsymbol{L} \cdot \boldsymbol{S}
\end{gather}
with the mass $m$ of a particle.
In this case, an energy eigenstate is characterized by the principal quantum number $n$, the orbital angular momentum $l$ of the large component, the total angular momentum $j$ and its $z$ component, $j_z$.
The first-order correction  coming from  $H_{\mathrm{SO}}$ to the unperturbed energy eigenvalue $E_{nlj}$ is thus the diagonal matrix element:
\begin{gather}
	\Delta E_{nlj} =
		\frac{\xi_{nl}}{2} \Bigg[ j(j+1) - l(l+1) - \frac{3}{4} \Bigg]
	,
\end{gather}
where
$\xi_{nl} \equiv ( 2m^2 c^2 )^{-1} \langle r^{-1} \diff V/ \diff r  \rangle_{nl}$.
The energy splitting between the states with a common $l$ is calculated as
\begin{gather}
	\Delta E_{nl}
		= \Delta E_{n l l + 1/2} -  \Delta E_{n l l - 1/2}
		= \xi_{nl} \Bigg( l + \frac{1}{2} \Bigg)
	.
\end{gather}
Our estimated values are as follows:
$\xi_{3 d} = 61$ meV for Fe, 
$\xi_{5 d} = 570$ meV for Pt, 
$\xi_{3 d} = 74$ meV for Co, 
$\xi_{4 d} = 240$ meV for Pd, 
$\xi_{3 d} = 48$ meV for Mn, 
$\xi_{3 p} = 11$ meV for Al, and
$\xi_{4 p} = 81$ meV for Ga.

It is known that the ordinary DFT functional does not contain terms responsible for the Hund's second rule,
which requires that not only the spin but also the orbital angular momentum of a system be maximized.
Jansen\cite{bib:2908} demonstrated that the energy functional must contain the term
which depend explicitly on the OAM for the description of the Hund's second rule for the accurate reproduction of the measured magnetism.
The orbital polarization correction (OPC) for a DFT calculation was introduced by Brooks\cite{bib:2458_10} employing the vector model\cite{bib:vector_model}.
This prescription has been applied not only to $5 f$ narrow-band compounds\cite{bib:2763} but also to $L1_0$ alloys\cite{bib:2458, bib:2608}.
There also exist, on the other hand, criticisms stating that 
the underestimation of OAM in DFT calculations using ordinary functionals such as LDA and GGA comes from different physics than that assumed in the OPC method\cite{bib:410,bib:2958_18,bib:2958}.
We do not take into account the OPC in the present study,
since our main purpose is to examine the validity and the reliability of the second-order perturbation formula derived above
by comparing it with the total energies obtained in self-consistent FR DFT calculations.

\subsubsection{MCA energies in DFT total-energy calculations}

We define the MCA energy of an $L1_0$ alloy as the difference in total energy between the spins of the transition metal atoms along the $[100]$ and the $[001]$ direction:
\begin{gather}
	E_\mathrm{MCA} \equiv E_{[100]} - E_{[001]}
	.
\end{gather}
A positive (negative) $E_\mathrm{MCA}$ means the magnetization easy axis along the $[001]$ (the $[100]$) direction.
In the present study, we calculate the MCA energies by calculating the differences in total energy between the self-consistent FR DFT calculations for differenet spin directions. 
The force theorem\cite{bib:Force_Theorem, bib:2196} cannot be used in our case since its mathematical validity is ensured only for a perturbative DFT calculation in which the charge density is not relaxed.

For each of the five $L1_0$ alloys,
we obtained the MCA energy in self-consistent FR DFT total-energy calculations
and show them in Table \ref{table_L10_MAE}, together with the results in the literature.
The reasonable agreement between our results and the previous results is seen.
The easy axes are in the $[001]$ directions for all the systems.
It is noticed that the Pt and the Pd atoms has the significant magnitudes of spins, though they are not magnetic elements.
Their spins originate in the hybridization between the $d$ orbitals at the magnetic and the nonmagnetic elements.
For each atom in of each system, the magnitude of the spin is found to be almost unchanged when its direction is changed,
while that of the OAM exhibits anisotropy for FePt, CoPt, and FePd.
One could expect that the MCA in these three alloys comes directly from the anisotropy of OAM.
MnAl and MnGa exhibit, however, the MCA energy larger than that of FePd despite their much weaker anisotropy of OAM.
These observations suggest that the origin of $L1_0$ alloys need to be examined in more detail.

\begin{table*}[tb]
\begin{center} 
\caption{
For each of the five $L1_0$ alloys comprised of the elements A and B (A = Fe, Co, or Mn, and B = Pt, Pd, Al, or Ga),
the calculated values of MCA energy in meV are tabulated.
The net OAM and the spin angular momentum in a.u. around each atom projected along the spin direction ($[100]$ or $[001]$) are also tabulated.
The $[100]$ and the $[001]$ directions correspond to the $x$ and the $z$ directions, respectively, in Fig. \ref{fig_L10_lattice}.
We tabulate the spin angular momenta multipled by $2$ since the literature provides not the spin angular momenta but the spin magnetic moments with a $g$ factor of $2$.
}
\label{table_L10_MAE}
\begin{tabular}{ccccccc}
\hline
System &  & MCA energy & $L_\mathrm{A}$ & $L_\mathrm{B}$ & $2 S_\mathrm{A}$ & $2 S_\mathrm{B}$ \\
\hline
FePt & Present work & $3.145$ &  &  &  &  \\
 & $[100]$ &  & $0.055$ & $0.057$ & $2.79$ & $0.37$ \\
 & $[001]$ &  & $0.060$ & $0.044$ & $2.79$ & $0.37$ \\
 & Daalderop \etal\cite{bib:2458} & $3.3$ & $0.08$ & $0.07$ & $2.91$ & $0.34$ \\
 & Sakuma\cite{bib:2224} & $2.8$ & $0.08$ & $0.05$ & $2.93$ & $0.33$ \\
 & Galanakis \etal\cite{bib:2625}  & $3.90$ & $0.07$ & $0.05$ & $2.88$ & $0.33$ \\
 & Ravindran \etal\cite{bib:2608} & $2.734$ &  &  &  &  \\
 & $[100]$ &  & $0.061$ & $0.055$ & $2.89$ & $0.355$ \\
 & $[001]$ &  & $0.067$ & $0.042$ & $2.89$ & $0.353$ \\
 & Burkert \etal\cite{bib:2600} & $2.84$ & $0.069, 0.078$ & $0.045, 0.043$ & $2.923, 2.937$ & $0.3615, 0.296$ \\
 & Lu \etal\cite{bib:2223} & $2.900$ &  &  &  &  \\
 &  &  &  &  &  &  \\
CoPt & Present work & $1.307$ &  &  &  &  \\
 & $[100]$ &  & $0.055$ & $0.078$ & $1.78$ & $0.40$ \\
 & $[001]$ &  & $0.089$ & $0.059$ & $1.77$ & $0.40$ \\
 & Daalderop \etal\cite{bib:2458} & $2.0$ & $0.12$ & $0.06$ & $1.86$ & $0.32$ \\
 & Sakuma\cite{bib:2224} & $1.5$ & $0.11$ & $0.07$ & $1.91$ & $0.38$ \\
 & Galanakis \etal\cite{bib:2625} & $2.20$ &  &  &  &  \\
 & $[100]$ &  & $0.06$ & $0.08$ & $1.74$ & $0.35$ \\
 & $[001]$ &  & $0.11$ & $0.06$ & $1.74$ & $0.35$ \\
 & Ravindran \etal\cite{bib:2608} & $1.052$ &  &  &  &  \\
 & $[100]$ &  & $0.057$ & $0.073$ & $1.809$ & $0.398$ \\
 & $[001]$ &  & $0.089$ & $0.056$ & $1.803$ & $0.394$ \\
 &  &  &  &  &  &  \\
FePd & Present work & $0.087$ &  &  &  &  \\
 & $[100]$ &  & $0.060$ & $0.030$ & $2.88$ & $0.38$ \\
 & $[001]$ &  & $0.070$ & $0.027$ & $2.88$ & $0.38$ \\
 & Galanakis \etal\cite{bib:2625} & $0.06$ &  &  &  &  \\
 & $[100]$ &  & $0.06$ & $0.03$ & $2.90$ & $0.35$ \\
 & $[001]$ &  & $0.07$ & $0.03$ & $2.90$ & $0.35$ \\
 &  &  &  &  &  &  \\
MnAl & Present work & $0.312$ &  &  &  &  \\
 & $[100]$ &  & $0.028$ & $0$ & $2.10$ & $-0.033$ \\
 & $[001]$ &  & $0.028$ & $-0.001$ & $2.10$ & $-0.033$ \\
 & Sakuma\cite{bib:2244} & $0.26$ & $0.059$ & $-0.003$ & $2.442$ & $-0.095$ \\
 &  &  &  &  &  &  \\
MnGa & Present work & $0.395$ &  &  &  &  \\
 & $[100]$ &  & $0.024$ & $0$ & $2.26$ & $-0.066$ \\
 & $[001]$ &  & $0.022$ & $0$ & $2.26$ & $-0.066$ \\
 & Sakuma\cite{bib:2245} & $0.42$ & $0.056$ & $0.005$ & $2.449$ & $-0.088$ \\
\hline
\end{tabular}
\end{center}
\end{table*}

\subsubsection{Analysis using the perturbation formula}

\begin{figure}[htbp]
\begin{center}
\includegraphics[keepaspectratio,width=8.5cm]{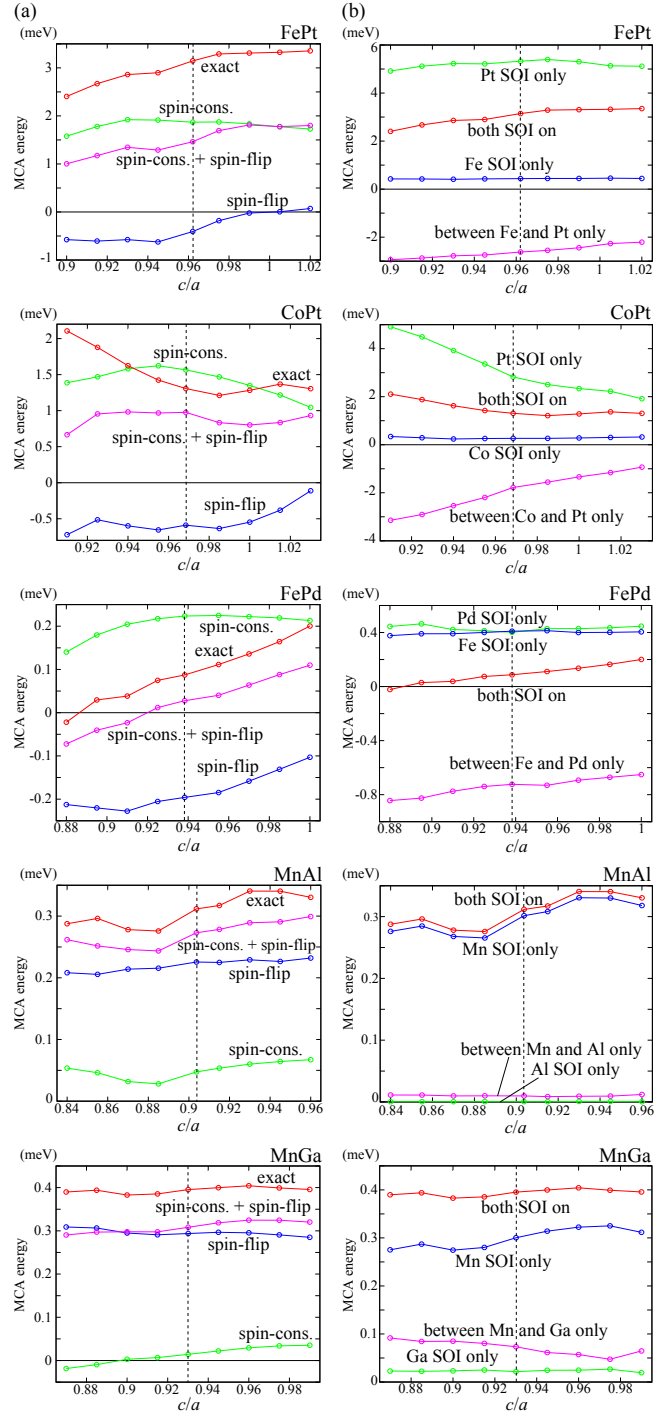}
\end{center}
\caption{
(Color online)
(a)
On the left panel for each of the five $L1_0$ alloys,
the exact MCA energy calculated as the difference in self-consistent FR DFT total energy between the electron spins in the $z$ and the $x$ directions
as a function of $c / a$ is plotted.
The contributions to the second-order perturbation formula calculated from the OAM matrices are also plotted.
The vertical dashed line corresponds to the experimental lattice constants for each system.
(b)
On the right panels, the exact MCA energies calculated with and without SOI of each element are plotted.
}
\label{fig_L10_MAE}
\end{figure}

For each of the five $L1_0$ alloys,
we obtained the MCA energy in the self-consistent FR DFT total-energy calculations as a function of $c / a$ is plotted in Fig. \ref{fig_L10_MAE} (a).
The contributions to the second-order perturbation formula, eq. (\ref{corr_formula}),
calculated from the OAM matrices, eq. (\ref{def_OAMmat}), are also plotted.
Though the variation in the Fermi level caused by the SOI is not taken into account in the formula,
it is seen for all the five systems that the formula reproduces qualitatively rather well the behavior of the exact values.
These results for the $L1_0$ alloys encourage us to use the second-order perturbation formula as a tractable tool for the analyses of MCA described in a self-consistent FR DFT calculation.

It is found for each of the systems in Fig. \ref{fig_L10_MAE} (a)
that the spin-conserving transitions contribute in favor of the spins in the $z$ direction.
The spin-flip transitions for FePt, CoPt, and FePd with their experimental lattice constants, however,
contribute in favor of the spins in the $x$ direction,
while those for MnAl and MnGa in favor of the spins in the $z$ direction.
The magnitudes of the spin-conserving contributions in FePt and CoPt are larger than the spin-flip contributions,
while the former contributions are much smaller in MnAl and MnGa.
The magnitudes of the spin-conserving and the spin-flip contributions are comparable in FePd.
The weak anisotropies of OAM in MnAl and MnGa (see Table \ref{table_L10_MAE}) are reflected in their small spin-conserving contributions,
which indicate that their MCA come mainly from the spin-flip transitions.

As $c/a$ increases,
the MCA energies of the systems except for CoPt tend to become higher.
Roughly speaking, these tendencies come from the reduction of dimensionality in the systems leading to the more localized valence electrons and the more effective SOI,
which reinforce their MCA.
It is observed in Fig. \ref{fig_L10_MAE} (a)
that the tendencies in FePt and FePd are from those of the spin-flip transitions,
while the tendencies in MnAl and MnGa are from those of the spin-conserving transitions.

Fig. \ref{fig_L10_MAE} (b) shows the MCA energies in the self-consistent FR DFT total-energy calculations as functions of $c / a$ with and without SOI of each element.
Those coming only from the interspecies transitions are calculated by using eq. (\ref{MAE_interspecies}) and also plotted.
For FePt and CoPt,
the contributions from the transitions between the magnetic atoms have only minor effects on the MCA energy.
In contrast, the MCA energy of MnAl and MnGa comes mainly from the transitions between the Mn atoms.
For FePd,
the transitions among the same species contribute to the MCA energy comparably to the interspecies transitions.
Is is found that the larger MCA of FePt and CoPt than FePd comes from the presence of the Pt atoms,
which has the quite large strength $\xi$ of SOI.

Since the spin-flip contributions in MnAl and MnGa are much larger than the spin-conserving contributions [see Fig. \ref{fig_L10_MAE} (a)],
we understand that their MCA energies come mainly from the spin-flip transitions occurring only around the Mn atoms.
It is the reason for the much smaller MCA energy of FePd than those of MnAl and MnGa despite the stronger anisotropy of OAM in FePd (see Table \ref{table_L10_MAE}).
Although stronger anisotropy of OAM, $\langle \boldsymbol{L} \rangle$, itself implies a larger contribution to the spin-conserving terms in the formula, eq. (\ref{corr_formula}),
it does not necessarily imply a larger contribution to the spin-flip terms, $\langle \boldsymbol{L} \cdot \boldsymbol{T} \rangle$.
These results indicate that the Bruno's formula is not sufficient for an analysis of the MCA of $L1_0$ alloys.

\subsubsection{OAM of FePt, CoPt, and FePd}

\begin{figure}[htbp]
\begin{center}
\includegraphics[keepaspectratio,width=8.5cm]{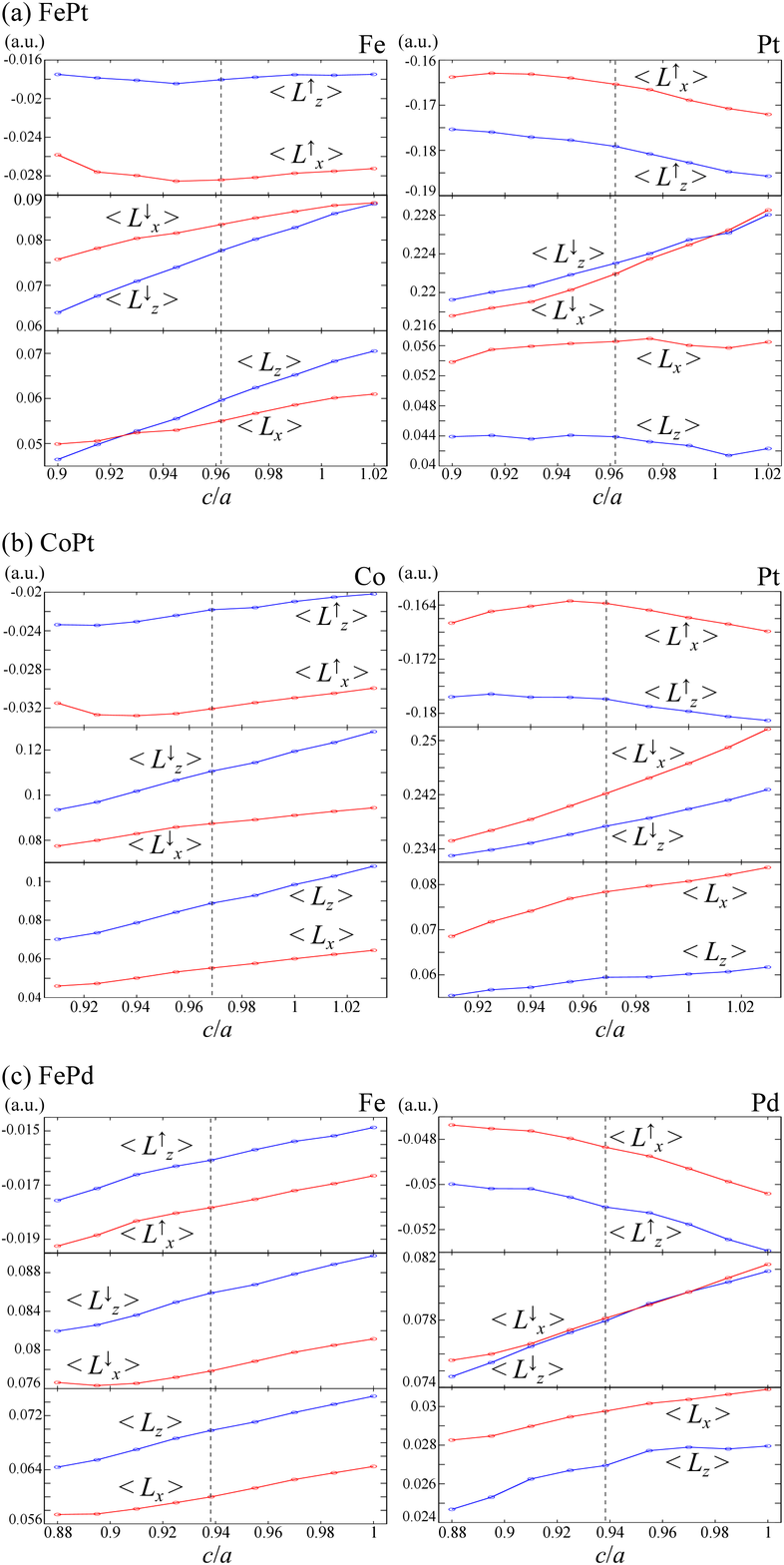}
\end{center}
\caption{
(Color online)
For each atom in (a) FePt, (b) CoPt, and (c) FePd with the electron spins in the $x$ and the $z$ directions,
the net OAM projected in the spin direction possessed by the spin-up and the spin-down electrons are plotted as functions of $c/a$.
The sums of the net OAM of the electrons of both spin directions are also plotted.
The dashed vertical lines correspond to the experimental lattice constants.
}
\label{fig_L10_OAM}
\end{figure}

For each atom in FePt, CoPt, and FePd with the electron spins in the $x$ and the $z$ directions,
the OAM projected in the spin direction possessed by the spin-up and the spin-down electrons are plotted in Fig. (\ref{fig_L10_OAM}) as functions of $c/a$.
In each atom $\mu$ in all the three systems,
the magnitude $\langle L_\mu^\uparrow \rangle$ of the OAM of the spin-up electrons is smaller than that of the spin-down electrons, $\langle L_\mu^\downarrow \rangle$.
It is because that the spin-up bands has the larger occupancy than the spin-down bands
and thus the spin-up electrons having opposite OAM among them have the stronger tendency to cancel their net OAM.
The OAM of the electrons in each spin direction for the three systems are in the opposite direction to the spin,
though their larger $\langle L_\mu^\downarrow \rangle$ dominate over their smaller $\langle L_\mu^\uparrow \rangle$ to give rise to the net OAM in the same direction as the spins.
The magnitudes  $\langle L_\mu \rangle$ of the net OAM at the atom $\mu$
become larger as $c/a$ increases, except for the Pt atom in FePt.
These tendencies of increase are the reflections of those of $\langle L_\mu^\downarrow \rangle$,
while $\langle L_\mu^\uparrow \rangle$ does not exhibit tendency of increase.
Since the spin-down states at the Fe or the Co atoms hybridize strongly with the states of both spin directions at the Pt or the Pd atoms\cite{bib:2608},
the OAM of the spin-down states are influenced sensitively by the variation in the distance between the atomic layers.
The OAM of the spin-up states, on the other hand, are influenced less sensitively by the interlayer distance than those of the spin-down states
and they do not necessarily exhibit tendency of increase as $c/a$ increases.

\begin{figure}[htbp]
\begin{center}
\includegraphics[keepaspectratio,width=5cm]{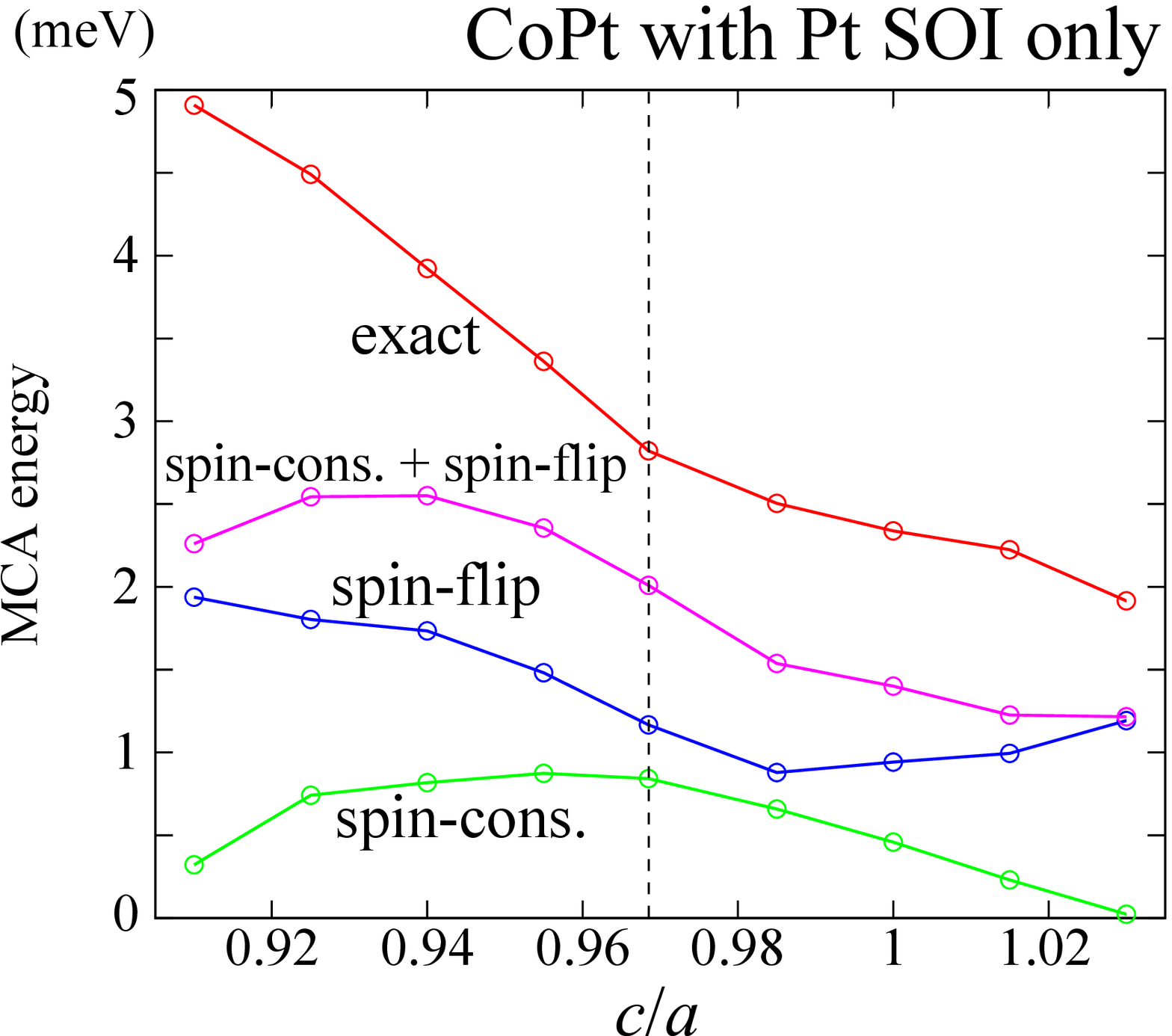}
\end{center}
\caption{
(Color online)
For CoPt with SOI only of Pt atoms,
the exact MCA energy calculated as the difference in self-consistent FR DFT total energy between the electron spins in the $z$ and the $x$ directions
as a function of $c / a$ is plotted.
The contributions to the second-order perturbation formula calculated from the OAM matrices are also plotted.
The vertical dashed line corresponds to the experimental lattice constants.
}
\label{Fig_L10_CoPt_MAE}
\end{figure}

As found above,
the exact MCA energy of CoPt for $0.91 < c/a < 0.98$ exhibits the tendency of decrease [see Fig. \ref{fig_L10_MAE} (a)], in contrast to FePt and FePd.
This tendency does not come from that in the spin-conserving contributions, which behave similarly in all the three systems.
The similarity of their behaviors is a direct consequence of that of the anisotropy of OAM in the three systems (see Fig. \ref{fig_L10_OAM}).
To identify the origin for this tendency,
the MCA energy of CoPt with the SOI only of the Pt atoms are plotted in Fig. \ref{Fig_L10_CoPt_MAE}.
It is seen in the figure that the exact MCA energy decreases monotonically as $c/a$ increases,
which should be regarded as the origin of the decrease in the MCA of CoPt seen in Fig. \ref{fig_L10_MAE} (a).
The spin-flip transitions occurring at the Pt atoms thus explain the decrease in the MCA energy of CoPt.

\section{Conclusions}

We derived succinctly a second-order perturbation formula for the correction to the energy eigenvalue of a many-body electronic system subject to SOI.
The energy correction was demonstrated to consist of three kinds of contributions:
the spin-conserving transitions of the spin-up electrons,
those of the spin-down electrons,
and the spin-flip transitions of the electrons of both spin directions.
The first two kinds of contributions are represented by the OAM acquired by the valence electrons via the SOI
The other kind of contributions was found to come from the quantum fluctuation effect.
In the limit of strong exchange interaction with completely filled majority spin bands, 
the formula derived is reduced to the well known Bruno's formula.
Since it uses only the wave functions of the perturbed system,
it serves as a tractable tool for the analyses of phenomena in which SOI plays important roles.
In particular, our formula provides a reliable way to capture essential physics of MCA.
By using our perturbation formula,
we examined the relativistic electronic structures of two examples, a $d$ orbital chain and $L1_{0}$ alloys.

The tight-binding calculations were performed for the $d$ orbital chain.
The appearance of OAM in the chain was clearly understood by using the parabolic-bands model and the exact expressions of the single-particle states.
The total energy as the sum of the energy eigenvalues of the single-particle states were found to be rather accurately reproduced by the formula,
though the formula does not take into account the variation in the Fermi level.

The first-principles calculations based on DFT were performed for the five $L1_0$ alloys
and their MCA energies were analyzed by using the OAM matrices.
We found that the formula reproduces qualitatively the behavior of the exact MCA energies of the alloys.
While the MCA of FePt, CoPt, and FePd was found to originate in the spin-conserving transitions,
that in MnAl and MnGa was found to originate in the spin-flip contributions.
For FePt, CoPt, and FePd,
the tendency of the MCA energy with the variation in $c/a$ was found to obey basically that of the spin-flip contributions.
These results indicate that not only the anisotropy of OAM $\langle \boldsymbol{L} \rangle$ itself,
but also that of spin-flip contribution, $\langle \boldsymbol{L} \cdot \boldsymbol{T} \rangle$,
must be taken into account for the understanding of the MCA of the $L1_0$ alloys.

Since the implementation of the OAM matrix is straightforward and it requires only the FR Bloch wave functions,
the second-order perturbation formula derived in the present study is tractable and useful for self-consistent FR DFT calculations.

\begin{acknowledgement}

The authors thank H. Kino, T. Shimazaki, and T. Nakajima for useful discussions.
Numerical calculation was partly carried out at the Supercomputer Center, ISSP, Univ. of Tokyo.
This work was supported by Grant-in-Aid for Scientific Research (No. 22104010 and No. 24540420),
Elements Strategy Initiative Center for Magnetic Materials under the outsourcing project of MEXT,
the Strategic Programs for Innovative Research (SPIRE), MEXT,
and the Computational Materials Science Initiative (CMSI), Japan.

\end{acknowledgement}

\appendix

\section{Exact Expressions of Fermi Level and OAM with SOI for Parabolic Bands}

In this Appendix, we derive the exact expressions of Fermi level and OAM with SOI turned on
for the bottoms of parabolic bands in one-, two-, and three-dimensional periodic systems.
The derivation of the expressions for the tops of parabolic bands will also be possible by considering the number of holes.

For each of the three cases below,
we assume that the two bands without SOI coincide with each other.
We further assume that all the unperturbed single-particle states in one band have the same OAM $m$ and those in the other band have the OAM with the same magnitude but in the opposite direction, $-m$.
We set the origin of energy to the bottom of the unperturbed bands.
For simplicity, we consider a case in which the band with the OAM $m$ undergoes a rigid shift $-b \xi$ as the perturbation, first-order in the strength $\xi$ of the SOI,
while the other band with $-m$ underdoes a rigid shift $b \xi$.
$b$ is a dimensionless positive constant.

\subsection{One-dimensional system}

Here we consider a one-dimensional periodic system.
For an energy $\varepsilon$ above the bottom of the unperturbed two bands,
the DOS is of the form $D_0 (\varepsilon) = 2 a / \sqrt{\varepsilon}$, where $a$ is a constant\cite{bib:Grosso_Parravicini}.
The number $n_e$ of electrons and the unperturbed Fermi level $\varepsilon_{\mathrm{F} 0}$ thus satisfy the relation
$n_e = 4 a \varepsilon_{\mathrm{F} 0}^{1/2} $.
The total energy of the unperturbed system is calculated as $E_0 = ( 4 a / 3) \varepsilon_{\mathrm{F} 0}^{3/2}$.
The DOS of the perturbed system is of the form $D (\varepsilon) = a / \sqrt{ \varepsilon - b \xi  } + a / \sqrt{ \varepsilon + b \xi  }$,
whose first term on the right hand side is for the states with the OAM $-m$ and the second term for $m$.

When $n_e$ is smaller than the critical value $n_\mathrm{c} \equiv 2 a ( 2 b \xi )^{1/2}$,
the band with OAM $-m$ is empty.
In such a case, the perturbed Fermi level is given by
\begin{gather}
	\varepsilon_\mathrm{F}
	=
		\left( \frac{ n_e}{2 a} \right)^2 - b \xi
	=
		\varepsilon_{\mathrm{F} 0}
		(  4 - \widetilde{\xi} )
	,
	\label{ef_para_1dim_large_xi}
\end{gather}
where $\widetilde{\xi} \equiv b \xi / \varepsilon_{\mathrm{F} 0} $ is the relative strength of the SOI with respect to the unperturbed Fermi level.
The net OAM then takes the saturated value
\begin{gather}
	\frac{\langle L \rangle}{ m n_e }
	=	\frac{1}{n_e } \int_{-\infty}^{\varepsilon_\mathrm{F}} \diff \varepsilon
		D ( \varepsilon )
	= 1
	,
	\label{oam_para_1dim_large_xi}
\end{gather}
independent of $\xi$.
The perturbed total energy is calculated as
\begin{gather}
	E
	=
		\int_{-\infty}^{\varepsilon_\mathrm{F}} \diff \varepsilon
		\varepsilon  D (\varepsilon)
	=
		E_0 (  4 - 3 \widetilde{\xi}  )
	.
\end{gather}
In the limit of $n_e \to n_\mathrm{c}$, the Fermi level converges as $\varepsilon_\mathrm{F} \to 2 \varepsilon_{\mathrm{F} 0}$.

When $n_e > n_\mathrm{c}$, on the other hand,
the perturbed Fermi level is determined so that
\begin{gather}
	n_e =	\int_{-\infty}^{\varepsilon_\mathrm{F}} \diff \varepsilon
		D (\varepsilon)
	=
		2 a [ ( \varepsilon_\mathrm{F} - b \xi  )^{1/2} + ( \varepsilon_\mathrm{F} + b \xi  )^{1/2} ]
	.
\end{gather}
By writing $\varepsilon_\mathrm{F} \equiv b \xi \cosh 2 s$ with $s > 0$,
this condition is written as $ \widetilde{n}_e =  e^s$,
where $\widetilde{n}_e \equiv n_e / n_\mathrm{c} $.
The perturbed Fermi level is thus given by
\begin{gather}
	\varepsilon_\mathrm{F}
	=
		\varepsilon_{\mathrm{F} 0}
		\left(
			1 + \frac{\widetilde{\xi^2}}{4}
		\right)
	,
	\label{ef_para_1dim_small_xi}
\end{gather}
The net OAM is hence calculated as
\begin{gather}
	\frac{\langle L \rangle}{m n_e}
	=	\frac{1}{n_e} \int_{-\infty}^{\varepsilon_\mathrm{F}} \diff \varepsilon
		\left( -\frac{a}{ \sqrt{ \varepsilon - b \xi  }} + \frac{a}{ \sqrt{ \varepsilon + b \xi } } \right)
	=
		\frac{\widetilde{\xi}}{2}
	,
	\label{oam_para_1dim_small_xi}
\end{gather}
which, in the limit of $n_e \to n_\mathrm{c}$ ($ \widetilde{\xi} \to 2$),
correctly converges to unity [see eq. (\ref{oam_para_1dim_large_xi})].
The Fermi level given by eq. (\ref{ef_para_1dim_small_xi}) also converges in this limit correctly as $\varepsilon_\mathrm{F} \to 2 \varepsilon_{\mathrm{F} 0}$ [see eq. (\ref{ef_para_1dim_large_xi})].
The perturbed total energy is calculated as
\begin{gather}
	E
	=
		\int_{-\infty}^{\varepsilon_\mathrm{F}} \diff \varepsilon
		\varepsilon D (\varepsilon)
	=
		E_0 \left(  1 - \frac{3}{4} \widetilde{\xi}^2  \right)
	.
\end{gather}

The Fermi level and the net OAM of the perturbed system as functions $\widetilde{\xi}$ are plotted in Fig. \ref{fig_dos_oam} (b).

\subsection{Two-dimensional system}

Here we consider a two-dimensional periodic system.
For an energy $\varepsilon$ above the bottom of the unperturbed two bands,
the DOS is of the form $D_0 (\varepsilon) = 2 a \theta ( \varepsilon )$, where $a$ is a constant for the step function\cite{bib:Grosso_Parravicini}.
The number $n_e$  of electrons and the unperturbed Fermi level $\varepsilon_{\mathrm{F} 0}$ thus satisfy the relation
$ n_e = 2 a \varepsilon_{\mathrm{F} 0}$.
The total energy of the unperturbed system is calculated as $E_0 = a \varepsilon_{\mathrm{F} 0}^2$.
The DOS of the perturbed system is of the form $D (\varepsilon) = a \theta ( \varepsilon - b \xi ) + a \theta ( \varepsilon + b \xi )$,
whose first term on the right hand side is for the states with the OAM $-m$ and the second term for $m$.

When $n_e$ is smaller than the critical value $n_\mathrm{c} \equiv a ( 2 b \xi )$,
the band with OAM $-m$ is empty.
In such a case, the perturbed Fermi level is given by
\begin{gather}
	\varepsilon_\mathrm{F}
	=
		\frac{ n_e}{a}  - b \xi
	=
		\varepsilon_{\mathrm{F} 0}
		(  2 - \widetilde{\xi} )
	,
	\label{ef_para_2dim_large_xi}
\end{gather}
where $\widetilde{\xi} \equiv b \xi / \varepsilon_{\mathrm{F} 0} $ is the relative strength of SOI with respect to the unperturbed Fermi level.
The net OAM then takes the saturated value
\begin{gather}
	\frac{\langle L \rangle}{ m n_e }
	=	\frac{1}{n_e } \int_{-\infty}^{\varepsilon_\mathrm{F}} \diff \varepsilon
		D ( \varepsilon )
	= 1
	,
	\label{oam_para_2dim_large_xi}
\end{gather}
independent of $\xi$.
The perturbed total energy is calculated as
\begin{gather}
	E
	=
		\int_{-\infty}^{\varepsilon_\mathrm{F}} \diff \varepsilon
		\varepsilon  D (\varepsilon) 
	=
		E_0 ( 2 - 2 \widetilde{\xi}  )
	.
\end{gather}
In the limit of $n_e \to n_\mathrm{c}$, the Fermi level converges as $\varepsilon_\mathrm{F} \to \varepsilon_{\mathrm{F} 0}$.

When $n_e > n_\mathrm{c}$, on the other hand,
the perturbed Fermi level is determined so that
\begin{gather}
	n_e =	\int_{-\infty}^{\varepsilon_\mathrm{F}} \diff \varepsilon
			D ( \varepsilon )
	=
		2 a \varepsilon_\mathrm{F}
	.
\end{gather}
The perturbed Fermi level is thus given by
\begin{gather}
	\varepsilon_\mathrm{F} = \frac{n_e}{2 a} = \varepsilon_{\mathrm{F} 0}
	\label{ef_para_2dim_small_xi}
	.
\end{gather}
The net OAM is calculated as
\begin{gather}
	\frac{\langle L \rangle}{m n_e}
	=	\frac{1}{n_e} \int_{-\infty}^{\varepsilon_\mathrm{F}} \diff \varepsilon
			[ -a \theta (\varepsilon - b \xi ) + a \theta (  \varepsilon + b \xi ) ]
	= \widetilde{\xi}
	,
	\label{oam_para_2dim_small_xi}
\end{gather}
which, in the limit of $n_e \to n_\mathrm{c}$ ($\widetilde{\xi} \to 1$),
correctly converges to unity [see eq. (\ref{oam_para_2dim_large_xi})].
The perturbed total energy is calculated as
\begin{gather}
	E
	=
		\int_{-\infty}^{\varepsilon_\mathrm{F}} \diff \varepsilon
		\varepsilon  D ( \varepsilon )
	=
		E_0 ( 1 - \widetilde{\xi}^2  )
\end{gather}

The Fermi level and the net OAM of the perturbed system as functions $\widetilde{\xi}$ are plotted in Fig. \ref{fig_dos_oam} (b).

\subsection{Three-dimensional system}

Here we consider a three-dimensional periodic system.
For an energy $\varepsilon$ above the bottom of the unperturbed two bands,
the DOS is of the form $D_0 (\varepsilon) = 2 a \sqrt{\varepsilon}$, where $a$ is a constant\cite{bib:Grosso_Parravicini}.
The number $n_e$  of electrons and the unperturbed Fermi level $\varepsilon_{\mathrm{F} 0}$ thus satisfy the relation
$ n_e = (4 a/ 3) \varepsilon_{\mathrm{F} 0}^{3/2} $.
The total energy of the unperturbed system is calculated as $E_0 = ( 4 a / 5) \varepsilon_{\mathrm{F} 0}^{5/2}$.
The DOS of the perturbed system is of the form $D (\varepsilon) = a \sqrt{ \varepsilon - b \xi  } + a \sqrt{ \varepsilon + b \xi  }$,
whose first term on the right hand side is for the states with the OAM $-m$ and the second term for $m$.

When $n_e$ is smaller than the critical value $n_\mathrm{c} \equiv (2 a / 3) ( 2 b \xi )^{3/2}$,
the band with OAM $-m$ is empty.
In such a case, the perturbed Fermi level is given by
\begin{gather}
	\varepsilon_\mathrm{F}
	=
		\left( \frac{3 n_e}{2 a} \right)^{2/3} - b \xi
	=
		\varepsilon_{\mathrm{F} 0}
		(  4^{1/3} - \widetilde{\xi}  )
	,
	\label{ef_para_3dim_large_xi}
\end{gather}
where $\widetilde{\xi} \equiv b \xi / \varepsilon_{\mathrm{F} 0} $ is the relative strength of SOI with respect to the unperturbed Fermi level.
The net OAM then takes the saturated value
\begin{gather}
	\frac{\langle L \rangle}{ m n_e }
	=	\frac{1}{n_e } \int_{-\infty}^{\varepsilon_\mathrm{F}} \diff \varepsilon
		D ( \varepsilon )
	= 1
	,
	\label{oam_para_3dim_large_xi}
\end{gather}
independent of $\xi$.
The perturbed total energy is calculated as
\begin{gather}
	E
	=
		\int_{-\infty}^{\varepsilon_\mathrm{F}} \diff \varepsilon
		\varepsilon D ( \varepsilon )
	=
		E_0 \left( 4^{1/3} - \frac{5}{3} \widetilde{\xi}  \right)
	.
\end{gather}
In the limit of $n_e \to n_\mathrm{c}$, the Fermi level converges as $\varepsilon_\mathrm{F} \to 2^{-1/3} \varepsilon_{\mathrm{F} 0}$.

When $n_e > n_\mathrm{c}$, on the other hand,
the perturbed Fermi level is determined so that
\begin{gather}
	n_e =	\int_{-\infty}^{\varepsilon_\mathrm{F}} \diff \varepsilon
		D ( \varepsilon )
	=
		a \frac{2}{3} [ ( \varepsilon_\mathrm{F} - b \xi  )^{3/2} + ( \varepsilon_\mathrm{F} + b \xi  )^{3/2} ]
	.
\end{gather}
By writing
\begin{gather}
	\varepsilon_\mathrm{F} \equiv b \xi \cosh 2 s
	\label{ef_para_3dim_small_xi}
\end{gather}
with $s > 0$,
this condition is written as 
$ n_e = ( n_\mathrm{c} / 4 ) ( e^{3 s} + 3 e^{-s} ) $.
Via a further variable transformation
$t \equiv e^s$,
the condition to be satisfied becomes
$ t^4 - 4 \widetilde{n}_e t + 3 = 0 $,
where $ \widetilde{n}_e \equiv n_e / n_\mathrm{c} $.
The only appropriate solution of this quartic equation for $s > 0$ is
$ t = u + \sqrt{  \widetilde{n}_e / u  - u^2 }$,
where
$ u \equiv \sqrt{ (h^2 + h^{-2}) / 2 }$
and
$ h \equiv
		[
			\widetilde{n}_e^2  + \sqrt{  \widetilde{n}_e^4 - 1 }
		]^{1/6}
$.
The perturbed Fermi level can be calculated by putting this solution into eq. (\ref{ef_para_3dim_small_xi}).
The net OAM is hence calculated as
\begin{gather}
	\frac{\langle L \rangle}{m n_e}
	 =	\frac{1}{n_e} \int_{-\infty}^{\varepsilon_\mathrm{F}} \diff \varepsilon
		( -a \sqrt{ \varepsilon - b \xi  } + a \sqrt{ \varepsilon + b \xi  } )
	\nonumber \\
	=
		\frac{1}{4 \widetilde{n}_e} ( t^{-3} + 3 t )
	,
	\label{oam_para_3dim_small_xi}
\end{gather}
which, in the limit of $n_e \to n_\mathrm{c}$ ($\widetilde{\xi} \to 2^{-1/3} $),
correctly converges to unity [see eq. (\ref{oam_para_3dim_large_xi})].
The Fermi level given by eq. (\ref{ef_para_3dim_small_xi}) also converges in this limit correctly as $\varepsilon_\mathrm{F} \to 2^{-1/3} \varepsilon_{\mathrm{F} 0}$ [see eq. (\ref{ef_para_3dim_large_xi})].
The perturbed total energy is calculated as
\begin{gather}
	E
	=
		\int_{-\infty}^{\varepsilon_\mathrm{F}} \diff \varepsilon
		\varepsilon  D ( \varepsilon )
	\nonumber \\
	=
		\frac{2 a}{15}
		\Bigg[
			  \sqrt{ \varepsilon_\mathrm{F} + b \xi} (3 \varepsilon_\mathrm{F}^2 + b \xi \varepsilon_\mathrm{F} - 2 b^2 \xi^2)
	\nonumber \\
			+ \sqrt{ \varepsilon_\mathrm{F} - b \xi} (3 \varepsilon_\mathrm{F}^2 - b \xi \varepsilon_\mathrm{F} - 2 b^2 \xi^2)
		\Bigg]	
	.
	\label{tote_para_3dim_small_xi}
\end{gather}

When the strength of SOI is small compared to the unperturbed Fermi level measured from the bottom of the bands,
the perturbed Fermi level for $n_e > n_\mathrm{c}$ is expressed, from eq. (\ref{ef_para_3dim_small_xi}), as
\begin{gather}
	\varepsilon_\mathrm{F}
	\approx
	\varepsilon_{\mathrm{F} 0}
		\left(
			1
			-\frac{\widetilde{\xi}^2}{4}
			-\frac{\widetilde{\xi}^4}{16}
		\right)
	.
	\label{ef_para_3dim_small_xi_approx}
\end{gather}
The net OAM is expressed, from eq. (\ref{oam_para_3dim_small_xi}), as
\begin{gather}
	\frac{\langle L \rangle}{m n_e}
	\approx
		\frac{3}{2} \widetilde{\xi}
		-\frac{1}{4} \widetilde{\xi}^3
	.
	\label{oam_para_3dim_small_xi_approx}
\end{gather}
The total energy is expressed, from eq. (\ref{tote_para_3dim_small_xi}), as
\begin{gather}
	E
	\approx
		E_0
		\left(
			1
			- \frac{5 \widetilde{\xi}^2}{4}
			+ \frac{5 \widetilde{\xi}^4}{48}
		\right)
	.
\end{gather}

The Fermi level and the net OAM of the perturbed system as functions $\widetilde{\xi}$ are plotted in Fig. \ref{fig_dos_oam} (b).

\section{Exact Expressions of Eigenvectors for $d$ Orbital Chain with Electron Spins along $z$ Axis}

In this Appendix, we provide the exact expressions of the energy eigenvalues and the eigenvectors for the $d$ orbital chain with electron spins along the $z$ axis ($\boldsymbol{n} = \boldsymbol{e}_z$).

Substituting $\theta = 0$ and $\phi = 0$ into the expressions of the spin wave function, eq. (\ref{two-comp_spinor}),
we obtain the basis functions in spin space as
\begin{gather}
	| \uparrow \rangle
	=
	\begin{pmatrix}
		1 \\
		0
	\end{pmatrix}
	, \,
	| \downarrow \rangle
	=
	\begin{pmatrix}
		0 \\
		-1
	\end{pmatrix}
	.
\end{gather}
We rearrange the basis functions for the chain as
$\{
d_{xy}^\uparrow, d_{x^2 - y^2}^\uparrow, d_{3 z^2 - r^2}^\uparrow, d_{yz}^\downarrow, d_{zx}^\downarrow,
d_{xy}^\downarrow, d_{x^2 - y^2}^\downarrow, d_{3 z^2 - r^2}^\downarrow, d_{yz}^\uparrow, d_{zx}^\uparrow
\}$.
The ten-dimensional Hamiltonian matrix,
which is represented by eq. (\ref{chain_H_mat}) for the old arrangement of the basis functions,
then becomes block diagonal consisting of two $5 \times 5$ matrices.
The Hamiltonian matrix for the former five basis functions reads
\begin{gather}
	H_1 (k)
	=
	\begin{pmatrix}
		2 t_\delta p - \Delta_\mathrm{ex}/2 & i \xi & 0 & -\xi /2  & i \xi /2   \\
		-i \xi & 2 t_\delta p - \Delta_\mathrm{ex}/2 & 0 & -i \xi /2  & -\xi /2  \\
		0 & 0 & 2 t_\sigma p - \Delta_\mathrm{ex}/2 & -i \xi \sqrt{3} /2  & \xi \sqrt{3} /2   \\
		-\xi /2   & i \xi /2   & i \xi \sqrt{3} /2  & 2 t_\pi p + \Delta_\mathrm{ex}/2 & -i \xi/2 \\
		i \xi /2   & -\xi /2   & \xi \sqrt{3} /2   & i \xi/2 & 2 t_\pi p + \Delta_\mathrm{ex}/2 \\
	\end{pmatrix}
	,
\end{gather}
while that for the latter reads
\begin{gather}
	H_2 (k)
	=
	\begin{pmatrix}
		2 t_\delta p + \Delta_\mathrm{ex}/2 & -i \xi & 0 & \xi /2   & i \xi /2   \\
		i \xi & 2 t_\delta p + \Delta_\mathrm{ex}/2 & 0 & -i \xi /2   & \xi /2   \\
		0 & 0 & 2 t_\sigma p + \Delta_\mathrm{ex}/2 & -i \xi \sqrt{3} /2  & -\xi \sqrt{3} /2   \\
		\xi /2  & i \xi /2  & i \xi \sqrt{3} /2   & 2 t_\pi p - \Delta_\mathrm{ex}/2 & i \xi/2 \\
		-i \xi /2  & \xi /2  & -\xi \sqrt{3} /2  & -i \xi/2 & 2 t_\pi p - \Delta_\mathrm{ex}/2 \\
	\end{pmatrix}
	,
\end{gather}
where $p \equiv \cos ka$.
The eigenvalue problem in this case has reduced to the two quintic equations.
These Hamiltonians are analytically diagonalizable.

We define the following dimensionless parameters:
\begin{gather}
	\eta_{(\pm)} \equiv \frac{\xi}{ 4 (t_\sigma - t_\pi ) p \pm 2 \Delta_\mathrm{ex} }
	, \\
	\gamma_{(\pm)} \equiv \frac{\xi}{ -4 (t_\delta - t_\pi ) p \pm 2 \Delta_\mathrm{ex} }
	.
\end{gather}
The exact energy eigenvalues of $H_1 (k)$ are then given by
\begin{gather}
	\varepsilon_{11}
	=
		(t_\pi + t_\sigma) p
		-\frac{\xi}{4} \left[  1 + \sqrt{ ( 1 + \eta_{(-)}^{-1} )^2 + 24 } \right]  
	, \\
	\varepsilon_{12}
	=
		(t_\pi + t_\sigma) p
		-\frac{\xi}{4} \left[  1 - \sqrt{ ( 1 + \eta_{(-)}^{-1} )^2 + 24 } \right]
	, \\
	\varepsilon_{13}
	=
		(t_\pi + t_\delta) p
		-\frac{\xi}{4} \left[ 1 + \sqrt{ ( 3 + \gamma_{(+)}^{-1} )^2 + 16 }   \right]
	, \\
	\varepsilon_{14}
	=
		(t_\pi + t_\delta) p
		-\frac{\xi}{4} \left[ 1 - \sqrt{ ( 3 + \gamma_{(+)}^{-1} )^2 + 16 }   \right]
	, \\
	\varepsilon_{15}
	=
		2  t_\delta  p - \frac{\Delta_\mathrm{ex}}{2}
		+ \xi
	,
\end{gather}
while those of $H_2 (k)$ are given by
\begin{gather}
	\varepsilon_{21}
	=
		(t_\pi + t_\sigma) p
		-\frac{\xi}{4} \left[  1 + \sqrt{ ( 1 + \eta_{(+)}^{-1} )^2 + 24 } \right]
	, \\
	\varepsilon_{22}
	=
		(t_\pi + t_\sigma) p
		-\frac{\xi}{4} \left[  1 - \sqrt{ ( 1 + \eta_{(+)}^{-1} )^2 + 24 } \right]
	, \\
	\varepsilon_{23}
	=
		(t_\pi + t_\delta) p
		-\frac{\xi}{4} \left[ 1 + \sqrt{ ( 3 + \gamma_{(-)}^{-1} )^2 + 16 }   \right]
	, \\
	\varepsilon_{24}
	=
		(t_\pi + t_\delta) p
		-\frac{\xi}{4} \left[ 1 - \sqrt{ ( 3 + \gamma_{(-)}^{-1} )^2 + 16 }   \right]
	, \\
	\varepsilon_{25}
	=
		2  t_\delta  p + \frac{\Delta_\mathrm{ex}}{2}
		+ \xi
	.
\end{gather}
We confirmed that these expressions give the correct energy eigenvalues by comparing them with those obtained via numerical diagonalization.
From the exact eigenvalues displayed above,
their expressions for $H_1 (k)$ correct up to second order in $\xi$ are calculated as
\begin{gather}
	\varepsilon_{11}
	\approx
		2 t_\pi p + \frac{\Delta_\mathrm{ex}}{2}
		- \frac{\xi}{2}
		- \frac{3  \xi^2}{ 4 ( t_\sigma- t_\pi ) p - 2 \Delta_\mathrm{ex} }
	, \\
	\varepsilon_{12}
	\approx
		2 t_\sigma p - \frac{\Delta_\mathrm{ex}}{2}
		+ \frac{3  \xi^2}{ 4 ( t_\sigma - t_\pi ) p - 2 \Delta_\mathrm{ex} }
	, \\
	\varepsilon_{13}
	\approx
		2 t_\delta p - \frac{\Delta_\mathrm{ex}}{2}
		- \xi - \frac{\xi^2}{ -2 ( t_\delta- t_\pi ) p + \Delta_\mathrm{ex} }
	, \\
	\varepsilon_{14}
	\approx
		2 t_\pi p + \frac{\Delta_\mathrm{ex}}{2}
		+ \frac{\xi}{2}
		+ \frac{\xi^2}{ -2 ( t_\delta- t_\pi ) p + \Delta_\mathrm{ex} }
	,
\end{gather}
while those of $H_2 (k)$ are given by
\begin{gather}
	\\
	\varepsilon_{21}
	\approx
		2 t_\pi p - \frac{\Delta_\mathrm{ex}}{2}
		-\frac{\xi}{2}
		- \frac{3  \xi^2}{ 4 ( t_\sigma- t_\pi ) p + 2 \Delta_\mathrm{ex} }
	, \\
	\varepsilon_{22}
	\approx
		2 t_\sigma p + \frac{\Delta_\mathrm{ex}}{2}
		+ \frac{3  \xi^2}{ 4 ( t_\sigma - t_\pi ) p + 2 \Delta_\mathrm{ex} }
	, \\
	\varepsilon_{23}
	\approx
		2 t_\delta p + \frac{\Delta_\mathrm{ex}}{2}
		- \xi
		- \frac{\xi^2}{ -2 ( t_\delta - t_\pi ) p - \Delta_\mathrm{ex} }
	, \\
	\varepsilon_{24}
	\approx
		2 t_\pi p - \frac{\Delta_\mathrm{ex}}{2}
		+ \frac{\xi}{2}
		+ \frac{\xi^2}{ -2 ( t_\delta - t_\pi ) p - \Delta_\mathrm{ex} }
	.
\end{gather}
These expressions allow one to find the correspondence between the perturbed energy eigenvalues and the unperturbed states [see Fig. \ref{fig_chain_bands} (b)]:
$\varepsilon_{13}$ and $\varepsilon_{15}$ from $\delta \uparrow$,
$\varepsilon_{23}$ and $\varepsilon_{25}$ from $\delta \downarrow$,
$\varepsilon_{21}$ and $\varepsilon_{24}$ from $\pi \uparrow$,
$\varepsilon_{11}$ and $\varepsilon_{14}$ from $\pi \downarrow$,
$\varepsilon_{12}$ from $\sigma \uparrow$,
$\varepsilon_{22}$ from $\sigma \downarrow$
states.

We define the following eight functions:
\begin{gather}
	f_\pm (q)
		\equiv
		\frac{ \sqrt{3} q [ \pm q \sqrt{ q^{-2} + 2 q^{-1} + 25 } -1 + 3 q  ]  }{ \pm q \sqrt{ q^{-2} + 2 q^{-1} + 25 } ( -1 + q) + 1 + 11 q^2}
	, \\
	g_\pm (q)
		\equiv
		\frac{\sqrt{3}}{12} [ \pm q \sqrt{ q^{-2} + 2 q^{-1} + 25 } - 1 - q ] / q
	, \\
	u_\pm (q)
		\equiv
		\frac{ 2 q [ \pm q \sqrt{ q^{-2} + 6 q^{-1} + 25 } +1 - q  ]  }{ \pm q \sqrt{ q^{-2} + 6 q^{-1} + 25 } ( 1 + q) + 1 + 4 q + 11 q^2}
	, \\
	v_\pm (q)
		\equiv
		\frac{1}{4} [ \pm q \sqrt{ q^{-2} + 6 q^{-1} + 25 } + 1 + 3 q ] / q
	.
\end{gather}
The normalized exact eigenvectors of $H_1 (k)$ corresponding to the energy eigenvalues provided above are then given by
\begin{gather}
	| \psi_{11} \rangle =
	\frac{1}{N_{11}}
	\begin{pmatrix}
		0 \\
		0 \\
		1 \\
		-i f_+ ( \eta_{(-)} ) \\
		g_- ( \eta_{(-)} )
	\end{pmatrix}
	, \\
	| \psi_{12} \rangle =
	\frac{1}{N_{12}}
	\begin{pmatrix}
		0 \\
		0 \\
		1 \\
		-i f_- ( \eta_{(-)} ) \\
		g_+ ( \eta_{(-)} )
	\end{pmatrix}
	, \\
	| \psi_{13} \rangle =
	\frac{1}{N_{13}}
	\begin{pmatrix}
		1 \\
		i \\
		0 \\
		u_+ ( \gamma_{(+)} ) \\
		-i v_- ( \gamma_{(+)} )
	\end{pmatrix}
	, \\
	| \psi_{14} \rangle =
	\frac{1}{N_{14}}
	\begin{pmatrix}
		1 \\
		i \\
		0 \\
		u_- ( \gamma_{(+)} ) \\
		-i v_+ ( \gamma_{(+)} )
	\end{pmatrix}
	, \\
	| \psi_{15} \rangle =
	\frac{1}{\sqrt{2}}
	\begin{pmatrix}
		1 \\
		-i \\
		0 \\
		0 \\
		0
	\end{pmatrix}
	,
\end{gather}
while those of $H_2 (k)$ are given by
\begin{gather}
	| \psi_{21} \rangle =
	\frac{1}{N_{21}}
	\begin{pmatrix}
		0 \\
		0 \\
		1 \\
		-i f_+ ( \eta_{(+)} ) \\
		-g_- ( \eta_{(+)} )
	\end{pmatrix}
	, \\
	| \psi_{22} \rangle =
	\frac{1}{N_{22}}
	\begin{pmatrix}
		0 \\
		0 \\
		1 \\
		-i f_- ( \eta_{(+)} ) \\
		-g_+ ( \eta_{(+)} )
	\end{pmatrix}
	, \\
	| \psi_{23} \rangle =
	\frac{1}{N_{23}}
	\begin{pmatrix}
		1 \\
		-i \\
		0 \\
		-u_+ ( \gamma_{(-)} ) \\
		-i v_- ( \gamma_{(-)} )
	\end{pmatrix}
	, \\
	| \psi_{24} \rangle =
	\frac{1}{N_{24}}
	\begin{pmatrix}
		1 \\
		-i \\
		0 \\
		-u_- ( \gamma_{(-)} ) \\
		-i v_+ ( \gamma_{(-)} )
	\end{pmatrix}
	, \\
	| \psi_{25} \rangle =
	\frac{1}{\sqrt{2}}
	\begin{pmatrix}
		1 \\
		i \\
		0 \\
		0 \\
		0
	\end{pmatrix}
	.
\end{gather}
$N_{ij}$'s ($i = 1,2, j = 1, \dots, 4$) are the normalization constants.
From the exact eigenvectors displayed above,
their expressions for $H_1 (k)$ correct up to second order in $\xi$ are calculated as
\begin{gather}
	| \psi_{11} \rangle
	\approx
	\frac{1}{\sqrt{2}}
	\begin{pmatrix}
		0 \\
		0 \\
		2 \sqrt{3} [ -\eta_{(-)} + \eta_{(-)}^2 ] \\
		i [ 1 - 3 \eta_{(-)}^2 ] \\
		1 - 3 \eta_{(-)}^2
	\end{pmatrix}
	, \\
	| \psi_{12} \rangle
	\approx
	\begin{pmatrix}
		0 \\
		0 \\
		1 - 3 \eta_{(-)}^2 \\
		i \sqrt{3} [ \eta_{(-)}  -  \eta_{(-)}^2 ] \\
		\sqrt{3} [ \eta_{(-)}  -  \eta_{(-)}^2 ]
	\end{pmatrix}
	, \\
	| \psi_{13} \rangle
	\approx
	\frac{1}{\sqrt{2}}
	\begin{pmatrix}
		1 - 2 \gamma_{(+)}^2 \\
		i [ 1 - 2 \gamma_{(+)}^2 ] \\
		0 \\
		2   [ \gamma_{(+)} - 3 \gamma_{(+)}^2 ] \\
		2 i [ \gamma_{(+)} - 3 \gamma_{(+)}^2 ] \\
	\end{pmatrix}
	, \\
	| \psi_{14} \rangle
	\approx
	\frac{1}{\sqrt{2}}
	\begin{pmatrix}
		2   [ \gamma_{(+)} - 3 \gamma_{(+)}^2 ] \\
		2 i [ \gamma_{(+)} - 3 \gamma_{(+)}^2 ] \\
		0 \\
		-1 + 2 \gamma_{(+)}^2 \\
		i [ -1 + 2 \gamma_{(+)}^2 ] \\
	\end{pmatrix}
	,
\end{gather}
while those for $H_2 (k)$ are calculated as
\begin{gather}
	| \psi_{21} \rangle
	\approx
	\frac{1}{\sqrt{2}}
	\begin{pmatrix}
		0 \\
		0 \\
		2 \sqrt{3} [ \eta_{(+)} - \eta_{(+)}^2 ] \\
		-i [ 1 - 3 \eta_{(+)}^2 ] \\
		1 - 3 \eta_{(+)}^2
	\end{pmatrix}
	, \\
	| \psi_{22} \rangle
	\approx
	\begin{pmatrix}
		0 \\
		0 \\
		1 - 3 \eta_{(+)}^2 \\
		i \sqrt{3} [ \eta_{(+)} - \eta_{(+)}^2  ] \\
		-\sqrt{3} [ \eta_{(+)} - \eta_{(+)}^2  ] \\
	\end{pmatrix}
	, \\
	| \psi_{23} \rangle
	\approx
	\frac{1}{\sqrt{2}}
	\begin{pmatrix}
		1 - 2 \gamma_{(-)}^2 \\
		-i [ 1 - 2 \gamma_{(-)}^2 ] \\
		0 \\
		-2   [ \gamma_{(-)} - 3 \gamma_{(-)}^2 ] \\
	 	 2 i [ \gamma_{(-)} - 3 \gamma_{(-)}^2 ] \\
	\end{pmatrix}
	, \\
	| \psi_{24} \rangle
	\approx
	\frac{1}{\sqrt{2}}
	\begin{pmatrix}
		2    [ \gamma_{(-)} - 3 \gamma_{(-)}^2 ] \\
		-2 i [ \gamma_{(-)} - 3 \gamma_{(-)}^2 ] \\
		0 \\
		1 - 2 \gamma_{(-)}^2 \\
		-i [ 1 - 2 \gamma_{(-)}^2 ] \\
	\end{pmatrix}
	.
\end{gather}

By using the expressions for the single-particle states provided above,
we obtain the expectation values of $L_z$ correct up to second order in $\xi$ as
\begin{gather}
	\langle \psi_{1 1 } | L_z | \psi_{1 1} \rangle
	\approx 1 - 6 \eta_{(-)}^2
	\label{chain_z_Lz_2nd_start}
	, \\
	\langle \psi_{1 2 } | L_z | \psi_{1 2} \rangle
	\approx
	6 \eta_{(-)}^2
	, \\
	\langle \psi_{1 3 } | L_z | \psi_{1 3} \rangle
	\approx
	-2 + 4 \gamma_{(+)}^2
	, \\
	\langle \psi_{1 4 } | L_z | \psi_{1 4} \rangle
	\approx
	-1 - 4 \gamma_{(+)}^2
	, \\
	\langle \psi_{1 5 } | L_z | \psi_{1 5} \rangle = 2 
	, \\
	\langle \psi_{2 1 } | L_z | \psi_{2 1} \rangle
	\approx -1 + 6 \eta_{(+)}^2
	, \\
	\langle \psi_{2 2 } | L_z | \psi_{2 2} \rangle
	\approx
	-6 \eta_{(+)}^2
	, \\
	\langle \psi_{2 3 } | L_z | \psi_{2 3} \rangle
	\approx
	2 - 4 \gamma_{(-)}^2
	, \\
	\langle \psi_{2 4 } | L_z | \psi_{2 4} \rangle
	\approx
	1 + 4 \gamma_{(-)}^2
	, \\
	\langle \psi_{2 5 } | L_z | \psi_{2 5} \rangle = -2 
	,
	\label{chain_z_Lz_2nd_end}
\end{gather}
among which none contains the first-order contribution.
It is also easily confirmed for all the single-particle states that the first-order contribution to the expectation values of $P_\uparrow L_z$ and $P_\downarrow L_z$ vanishes separately.
On the other hand, the expectation values of $\boldsymbol{L} \cdot \boldsymbol{T}$ correct up to second order in $\xi$ are calculated as
\begin{gather}
	\langle \psi_{1 1 } | \boldsymbol{L} \cdot \boldsymbol{T} | \psi_{1 1} \rangle
	\approx - 6 \eta_{(-)} + 6 \eta_{(-)}^2
	\label{chain_z_LT_2nd_start}
	, \\
	\langle \psi_{1 2 } | \boldsymbol{L} \cdot \boldsymbol{T} | \psi_{1 2} \rangle
	\approx 6 \eta_{(-)} - 6 \eta_{(-)}^2
	, \\
	\langle \psi_{1 3 } | \boldsymbol{L} \cdot \boldsymbol{T} | \psi_{1 3} \rangle
	\approx
	-4 \gamma_{(+)} + 12 \gamma_{(+)}^2
	, \\
	\langle \psi_{1 4 } | \boldsymbol{L} \cdot \boldsymbol{T} | \psi_{1 4} \rangle
	\approx
	4 \gamma_{(+)} - 12 \gamma_{(+)}^2
	, \\
	\langle \psi_{1 5 } | \boldsymbol{L} \cdot \boldsymbol{T} | \psi_{1 5} \rangle = 0 
	, \\
	\langle \psi_{2 1 } | \boldsymbol{L} \cdot \boldsymbol{T} | \psi_{2 1} \rangle
	\approx -6 \eta_{(+)} + 6 \eta_{(+)}^2
	, \\
	\langle \psi_{2 2 } | \boldsymbol{L} \cdot \boldsymbol{T} | \psi_{2 2} \rangle
	\approx 6 \eta_{(+)} - 6 \eta_{(+)}^2
	, \\
	\langle \psi_{2 3 } | \boldsymbol{L} \cdot \boldsymbol{T} | \psi_{2 3} \rangle
	\approx
	-4 \gamma_{(-)} + 12 \gamma_{(-)}^2
	, \\
	\langle \psi_{2 4 } | \boldsymbol{L} \cdot \boldsymbol{T} | \psi_{2 4} \rangle
	\approx
	4 \gamma_{(-)} - 12 \gamma_{(-)}^2
	, \\
	\langle \psi_{2 5 } | \boldsymbol{L} \cdot \boldsymbol{T} | \psi_{2 5} \rangle = 0
	,
	\label{chain_z_LT_2nd_end}
\end{gather}
which can contain the first-order contributions.

\section{Penalty Functional for DFT Calculations}

\subsection{Definition and Expressions}

The formulation of spin-constrained variational problems for the minimization of total energy in DFT calculations has been done in the literature.\cite{bib:penalty}
In this Appendix, we describe the explicit expressions of penalty functional for detailed specification of the spins of individual atoms.

The spin of the atom $\mu$ in a periodic system is evaluated as the sum of the contributions from the occupied Bloch states:
\begin{gather}
	\boldsymbol{S}_\mu = \sum_{m, \boldsymbol{k}}^{\mathrm{occ.}} \langle \psi_{m \boldsymbol{k}} | P_\mu \boldsymbol{S} P_\mu | \psi_{m \boldsymbol{k}} \rangle
	,
\end{gather}
where $\boldsymbol{k}$ is the wave vector and $m$ is the band index for a two-component Bloch state $| \psi_{m \boldsymbol{k}} \rangle$.

For detailed specification of the directions and/or magnitudes of the spins of individual atoms,
we define the penalty functional consisting of three parts as
$P \equiv P_{\mathrm{dir}} + P_{\mathrm{mag}} + P_{\mathrm{cone}}$,
where
\begin{gather}
	P_{\mathrm{dir}} \equiv
		A_{\mathrm{dir}} \sum_\mu
		\Bigg( \cos^{-1} \frac{\boldsymbol{S}_\mu \cdot \boldsymbol{d}_{0 \mu}}{S_\mu} - \delta_\mu \Bigg)^2
		\cdot
		\nonumber \\
		\cdot
		\theta( \cos \delta_\mu - \boldsymbol{S}_\mu \cdot \boldsymbol{d}_{0 \mu} / S_\mu )
	,
	\label{pfunc_dir}
	\\
	P_{\mathrm{mag}} \equiv
		A_{\mathrm{mag}} \sum_\mu ( | S_\mu - M_{0 \mu} | - \Delta_\mu )^2
		\cdot
		\nonumber \\
		\cdot
		\theta(| S_\mu - M_{0 \mu} | - \Delta_\mu)
	,
	\label{pfunc_mag}
	\\
	P_{\mathrm{cone}} \equiv
		A_{\mathrm{cone}} \sum_\mu
		\Bigg( \cos^{-1} \frac{\boldsymbol{S}_\mu \cdot \boldsymbol{c}_{0 \mu}}{S_\mu} - \theta_\mu \Bigg)^2
	\label{pfunc_cone}
	.
\end{gather}
$A_{\mathrm{dir}}, A_{\mathrm{mag}}$, and $A_{\mathrm{cone}}$ are positive constants.
$\theta$ is the step function.
$P_{\mathrm{dir}}$ is used for fixing the directions of the spins.
If the direction of $\boldsymbol{S}_\mu$ deviates from $\boldsymbol{d}_{0 \mu}$ by an angle larger than $\delta_\mu$,
$P_{\mathrm{dir}}$ has a positive value.
$P_{\mathrm{mag}}$ is used for fixing the magnitudes of the spins.
If the magnitude of $\boldsymbol{S}_\mu$ deviates from $M_{0 \mu}$ by a value larger than $\Delta_\mu$,
$P_{\mathrm{mag}}$ has a positive value.
$P_{\mathrm{cone}}$ is used for forcing the spins to be on the cones.
If $\boldsymbol{S}_\mu$ deviates from the cone whose axis is $\boldsymbol{c}_{0 \mu}$,
$P_{\mathrm{cone}}$ has a positive value.

The generalized energy functional to be minimized in this case is thus
$\widetilde{E} \equiv E + P$, where $E$ is the ordinary energy functional.
The penalty functional acts as the constraint for the energy minimization procedure in a DFT calculation.
It is noted, however, that the configuration of the spins does not necessarily minimize $P$ when the SCF calculation is converged,
since the energy minimization procedure minimizes not $P$, but $E + P$.

The equation to be solved is obtained from the stationarity condition of the generalized energy functional $\widetilde{E}$,
\begin{gather}
	0 = \frac{\delta E }{ \delta \langle \psi_{m \boldsymbol{k}} |}
		+ \frac{ \delta P}{\delta \langle \psi_{m \boldsymbol{k}} |}
	,
\end{gather}
where the first term on the right-hand side leads to the ordinary Kohn-Sham Hamiltonian for $| \psi_{m \boldsymbol{k}} \rangle$.
If we write the variation of the penalty functional as
\begin{gather}
	\frac{ \delta P}{\delta \langle \psi_{m \boldsymbol{k}} |}
	= \sum_\mu \frac{\partial P}{\partial  \boldsymbol{S}_\mu }
		\cdot \frac{ \delta  \boldsymbol{S}_\mu}{\delta \langle \psi_{m \boldsymbol{k}} |}
	\nonumber \\
	= \sum_\mu 
		 P_\mu \boldsymbol{B}_\mu^{\mathrm{pen}} \cdot \boldsymbol{S} P_\mu | \psi_{m \boldsymbol{k}} \rangle
	\equiv \sum_\mu 
		 P_\mu H_\mu^{\mathrm{pen}} P_\mu | \psi_{m \boldsymbol{k}} \rangle
	,
	\label{penvar}
\end{gather}
$\boldsymbol{B}_\mu^{\mathrm{pen}} \equiv \boldsymbol{B}_\mu^{\mathrm{dir}} + \boldsymbol{B}_\mu^{\mathrm{mag}} + \boldsymbol{B}_\mu^{\mathrm{cone}}$
can be interpreted as the effective magnetic field acting on the atom $\mu$ for fixing its spin.
From eqs. (\ref{pfunc_dir})-(\ref{pfunc_cone}),
the expressions for $\boldsymbol{B}_\mu^{\mathrm{pen}}$ are given by
\begin{gather}
	\boldsymbol{B}_{\mu i}^{\mathrm{dir}}
	= 2 A_{\mathrm{dir}} \theta( \cos \delta_\mu - d_\mu)
		\frac{\cos^{-1} d_\mu - \delta_\mu}{\sqrt{1 - d_\mu^2}}
		\cdot
		\nonumber \\
		\cdot
		\frac{1}{S_\mu} \Bigg( d_\mu \frac{\boldsymbol{S}_\mu}{S_\mu} - \boldsymbol{d}_{0 \mu}\Bigg)
	\\
	\boldsymbol{B}^{\mathrm{mag}}_{\mu} =
		2A_{\mathrm{mag}} [ S_\mu - M_{0 \mu} - \mathrm{sgn} (S_\mu - M_{0 \mu}) \Delta_\mu ]
		\cdot
		\nonumber \\
		\cdot
		\theta(| S_\mu - M_{0 \mu} | - \Delta_\mu)
		\frac{\boldsymbol{S}_\mu}{S_\mu}
	\\
	\boldsymbol{B}_{\mu}^{\mathrm{cone}}
	= 2 A_{\mathrm{cone}} 
		\frac{\cos^{-1} c_\mu - \theta_\mu}{\sqrt{1 - c_\mu^2}}
		\frac{1}{S_\mu} \Bigg( c_\mu \frac{\boldsymbol{S}_\mu}{S_\mu} - \boldsymbol{c}_{0 \mu}\Bigg)
	,
\end{gather}
where $d_\mu \equiv \boldsymbol{S}_\mu \cdot \boldsymbol{d}_{0 \mu} / S_\mu, c_\mu \equiv \boldsymbol{S}_\mu \cdot \boldsymbol{c}_{0 \mu} / S_\mu$.

\subsection{Implementation for PAW Method}

Within the PAW formalism\cite{bib:PAW}, an AE wave function and its corresponding PS wave function are
related via the transformation operator $T$ as $| \psi^{\mathrm{AE}} \rangle = T | \psi^{\mathrm{PS}} \rangle$.
The physical quantity represented by an AE operator $O^{\mathrm{AE}}$ is evaluated using the expectation value of the PS operator defined as
\begin{gather}
	O^{\mathrm{PS}} \equiv T^\dagger O^{\mathrm{AE}}  T
	= O^{\mathrm{AE}}
	\nonumber \\
	+ \sum_{\mu, i,j} 
	[ \langle \phi^{\mathrm{AE}}_{\mu i} | O^{\mathrm{AE}}  | \phi^{\mathrm{AE}}_{\mu j} \rangle -
	\langle \phi^{\mathrm{PS}}_{\mu i} | O^{\mathrm{AE}}  | \phi^{\mathrm{PS}}_{\mu j} \rangle ]
	| \beta_{\mu i} \rangle \langle \beta_{\mu j} |
	,
	\label{def_PS_operator}
\end{gather}
where $| \phi^{\mathrm{AE}}_{\mu i} \rangle$ and $| \phi^{\mathrm{PS}}_{\mu i} \rangle$ are the $i$th AE and PS atomic orbitals of the $\mu$th atom, respectively.
$| \beta_{\mu i} \rangle$ is the corresponding projector.
It is noted that the atomic orbitals and the projectors are two-component in our fully relativistic calculations.

The constrained minimization procedure of the total energy using a plane-wave basis set needs the matrix elements
$\langle \boldsymbol{k} + \boldsymbol{G}, \tau | H_\mu^\mathrm{pen} | \psi_{m \boldsymbol{k} } \rangle$ of the penalty Hamiltonian,
where
\begin{gather}
	| \boldsymbol{k} + \boldsymbol{G}, \alpha \rangle \equiv
	\begin{pmatrix}
		| \boldsymbol{k} + \boldsymbol{G} \rangle \\
		0
	\end{pmatrix}
	,
	| \boldsymbol{k} + \boldsymbol{G}, \beta \rangle \equiv
	\begin{pmatrix}
		0 \\
		| \boldsymbol{k} + \boldsymbol{G} \rangle
	\end{pmatrix}
\end{gather}
are the two-component PS plane waves.
From eqs. (\ref{penvar}) and (\ref{def_PS_operator}), we obtain
\begin{gather}
	\langle \boldsymbol{k} + \boldsymbol{G}, \tau |  
		( P_\mu H_\mu^{\mathrm{pen}} P_\mu )^{\mathrm{PS}} | \psi_{m \boldsymbol{k}}  \rangle
	\\ \nonumber
	= \sum_{\tau'} \langle \boldsymbol{k} + \boldsymbol{G} | Q_{\mu \tau \tau'} | \psi_{m \boldsymbol{k} \tau'} \rangle
	\\ \nonumber
		+  \sum_{i,j} L_{\mu ij} 
		\langle \boldsymbol{k} + \boldsymbol{G} | \beta_{\mu i \tau} \rangle
		\sum_{\tau'} \langle \beta_{\mu j \tau'} | \psi_{m \boldsymbol{k} \tau'} \rangle
	,
	\label{matel_Hpen}
\end{gather}
where
\begin{gather}
	Q_{\mu \tau \tau'} \equiv H^{\mathrm{pen}}_{\mu \tau \tau'} P_\mu 
\end{gather}
and
\begin{gather}
	L_{\mu ij} = \sum_{\tau, \tau'}
		H^{\mathrm{pen}}_{\mu \tau \tau'}
		\sqrt{4 \pi} q_{ij}^{\mu, 00, \tau \tau'}
	,
	\\
	q_{ij}^{\mu, lm, \tau \tau'}
	\equiv 
	\langle \phi^{\mathrm{AE}}_{\mu i \tau} | Y_{lm} | \phi^{\mathrm{AE}}_{\mu j \tau'} \rangle -
	\langle \phi^{\mathrm{PS}}_{\mu i \tau} | Y_{lm} | \phi^{\mathrm{PS}}_{\mu j \tau'} \rangle
	.
\end{gather}
$Y_{lm}$ is the spherical harmonics.
It is obvious from eq. (\ref{matel_Hpen}) that
the introduction of the penalty functional is realized only by replacing the local potential of the atom $\mu$ as
\begin{gather}
	V^{\mathrm{loc}}_{\mu \tau \tau'} (\boldsymbol{r})
	\rightarrow V^{\mathrm{loc}}_{\mu \tau \tau'} (\boldsymbol{r}) + Q_{\mu \tau \tau'} (\boldsymbol{r})
\end{gather}
and by replacing the coefficients of the nonlocal potential 
$V_\mu^{\mathrm{nonl}} = \sum_{i,j} D_{\mu ij} |\beta_{\mu i} \rangle \langle \beta_{\mu j} |$
as
\begin{gather}
	D_{\mu ij} \rightarrow D_{\mu ij} + L_{\mu ij}
	.
\end{gather}

\section{Calculation of OAM within PAW Method}

Here we describe the implementation of the OAM of each atom in a periodic system within the PAW method using a plane-wave basis set.

The OAM of the atom $\mu$ in a periodic system is evaluated as the sum of the contributions from the occupied Bloch states:
\begin{gather}
	\langle \boldsymbol{L}_\mu \rangle = \sum_{m, \boldsymbol{k}}^{\mathrm{occ.}} \langle \psi_{m \boldsymbol{k}} | P_\mu \boldsymbol{L}_{\mu} P_\mu | \psi_{m \boldsymbol{k}} \rangle
	,
\end{gather}
where $\boldsymbol{k}$ is the wave vector and $m$ is the band index for a two-component PS Bloch state $| \psi_{m \boldsymbol{k}} \rangle$.
$\boldsymbol{L}_{\mu} = (\boldsymbol{r} - \boldsymbol{R}_\mu) \times \boldsymbol{p}$ is the OAM operator effective only in the vicinity of the atom $\mu$, located at $\boldsymbol{R}_\mu$.
Within the PAW formalism\cite{bib:PAW}, 
a physical quantity is calculated by using the PS operator, defined via the relation eq. (\ref{def_PS_operator}), and the PS wave functions, as stated above.
The OAM of the atom $\mu$ is thus calculated as
\begin{gather}
	\langle \boldsymbol{L}_\mu \rangle
		= \sum_{m, \boldsymbol{k}}^{\mathrm{occ.}}
		\Bigg[
			\langle \psi_{m \boldsymbol{k}} | P_\mu \boldsymbol{L}_{\mu} P_\mu | \psi_{m \boldsymbol{k}} \rangle
			+ 
		\nonumber \\
			\sum_{i,j} 
			[ \langle \phi^{\mathrm{AE}}_{\mu i} | \boldsymbol{L} | \phi^{\mathrm{AE}}_{\mu j} \rangle -
			\langle \phi^{\mathrm{PS}}_{\mu i} | \boldsymbol{L} | \phi^{\mathrm{PS}}_{\mu j} \rangle ]
			\langle \psi_{m \boldsymbol{k}} | \beta_{\mu i} \rangle \langle \beta_{\mu j} | \psi_{m \boldsymbol{k}} \rangle
		\Bigg]
	.
	\label{L_mu_PAW}
\end{gather}
The first term in the summation on the right hand side above is written as
\begin{gather}
	\sum_{m, \boldsymbol{k}}^{\mathrm{occ.}}
		\langle \psi_{m \boldsymbol{k}} | P_\mu \boldsymbol{L}_{\mu} P_\mu | \psi_{m \boldsymbol{k}} \rangle
	\nonumber \\
	=
	\sum_{m, \boldsymbol{k}}^{\mathrm{occ.}}
		\int_{\mu} \diff^3 r \, \psi_{m \boldsymbol{k}} (\boldsymbol{r})^\dagger ( \boldsymbol{r} - \boldsymbol{R}_{\mu} ) \times \boldsymbol{p} \psi_{m \boldsymbol{k}} (\boldsymbol{r})
	\label{pw_plp}
	,
\end{gather}
where the integral is taken over the sphere of the ion radius $r_\mu$ centered at $\boldsymbol{R}_\mu$.
By defining a cell-periodic function
\begin{gather}
	\boldsymbol{P}_{m \boldsymbol{k}} (\boldsymbol{r}) \equiv \psi_{m \boldsymbol{k}} (\boldsymbol{r})^\dagger \boldsymbol{p} \psi_{m \boldsymbol{k}} (\boldsymbol{r})
	,
\end{gather}
we rewrite the right hand side of eq. (\ref{pw_plp}) as
\begin{gather}
	\sum_{m, \boldsymbol{k}}^{\mathrm{occ.}}
		\int_{\mu} \diff^3 r \, ( \boldsymbol{r} - \boldsymbol{R}_{\mu} ) \times \boldsymbol{P}_{m \boldsymbol{k}} (\boldsymbol{r})
	=
	- \sum_{m, \boldsymbol{k}}^{\mathrm{occ.}}
		\sum_{\boldsymbol{G}} e^{i \boldsymbol{G} \cdot \boldsymbol{R}_\mu}
		\cdot
	\nonumber \\
		\cdot
		\boldsymbol{P}_{m \boldsymbol{k} + \boldsymbol{G}} \times \Bigg( -i \frac{\partial}{\partial \boldsymbol{G}} \Bigg) 
		\int_0^{r_\mu} r^2 \diff r \int \diff \Omega  \,   e^{i \boldsymbol{G} \cdot \boldsymbol{r}}
	,
\end{gather}
where $\boldsymbol{G}$ is a reciprocal lattice vector and $\boldsymbol{P}_{m \boldsymbol{k} + \boldsymbol{G}}$ is the Fourier coefficient of $\boldsymbol{P}_{m \boldsymbol{k}} (\boldsymbol{r})$.
The integral on the right hand side above is performed as
\begin{gather}
	\frac{\partial}{\partial \boldsymbol{G}}
		\int_0^{r_\mu} r^2 \diff r \int \diff \Omega  \,  e^{i \boldsymbol{G} \cdot \boldsymbol{r}}
	\nonumber \\
	= 4 \pi r_\mu^4 \frac{\boldsymbol{G}_\mu}{G_\mu}
		\frac{(G_\mu^2 - 3) \sin G_\mu + 3 G_\mu \cos G_\mu}{G_\mu^4}
	,
\end{gather}
where $\boldsymbol{G}_\mu \equiv \boldsymbol{G} r_\mu$.
Equation (\ref{pw_plp}) is thus written as
\begin{gather}
	\sum_{m, \boldsymbol{k}}^{\mathrm{occ.}}
		\langle \psi_{m \boldsymbol{k}} | P_\mu \boldsymbol{L}_{\mu} P_\mu | \psi_{m \boldsymbol{k}} \rangle
	=
	 i 4 \pi r_\mu^4 \sum_{m, \boldsymbol{k}}^{\mathrm{occ.}}
		\sum_{\boldsymbol{G}} e^{i \boldsymbol{G} \cdot \boldsymbol{R}_\mu}
	\cdot
	\nonumber \\
	\cdot
		\boldsymbol{P}_{m \boldsymbol{k} + \boldsymbol{G}} \times 
	 \frac{\boldsymbol{G}_\mu}{G_\mu}
		\frac{(G_\mu^2 - 3) \sin G_\mu + 3 G_\mu \cos G_\mu}{G_\mu^4}
	.
\end{gather}
By using this expression in eq. (\ref{L_mu_PAW}),
one can evaluate the OAM of the atom $\mu$ straightforwardly in a PAW calculation using a plane-wave basis set.

The OAM matrix, defined in eq. (\ref{def_OAMmat}), can also be evaluated in a manner similar to that described above.

\end{document}